\documentclass[letterpaper,twoside,english,final]{IEEEtran}
\usepackage[T1]{fontenc}
\usepackage[latin9]{inputenc}
\usepackage{amsmath}
\usepackage{graphicx}
\usepackage{amssymb}

\makeatletter

\providecommand{\tabularnewline}{\\}

\newcounter{mytempeqncnt}

\author{Tarik~Ait-Idir,~\IEEEmembership{Member,~IEEE, }and Samir~Saoudi,~\IEEEmembership{Member,~IEEE}%
\thanks{Paper approved by A. Lozano, the Editor for Wireless Network Access and Performance of the IEEE Communications Society. Manuscript received July 7, 2008; revised May 27, 2009. This work was partly supported by Maroc Telecom under contract number 105 10005462.06/PI. This paper was presented in part at the IEEE Wireless Communications and Networking Conference, Las Vegas, NV, March-April 2008, and in part at the IEEE International Workshop on Signal Processing and Applications, Sharjah, UAE, March 2008.}%
\thanks{T. Ait-Idir is with the Communication Systems Department, INPT, Madinat Al-Irfane, Rabat, Morocco. He is also with Institut Telecom / Telecom Bretegne/LabSticc, Brest, France (email: aitidir@ieee.org).}%
\thanks{ S. Saoudi is with Institut Telecom / Telecom Bretegne/LabSticc, Brest, France. He is also with Université Européenne de Bretagne.}}%
\vspace{1cm}

\makeatother

\usepackage{babel}

\begin{document}

\title{Turbo Packet Combining Strategies for the MIMO-ISI ARQ Channel}
\maketitle
\begin{abstract}
This paper addresses the issue of efficient turbo packet combining
techniques for coded transmission with a Chase-type automatic repeat
request (ARQ) protocol operating over a multiple-input--multiple-output
(MIMO) channel with intersymbol interference (ISI). First of all,
we investigate the outage probability and the outage-based power loss
of the MIMO-ISI ARQ channel when optimal \emph{maximum a posteriori}
(MAP) turbo packet combining is used at the receiver. We show that
the ARQ delay (i.e., the maximum number of ARQ rounds) does not completely
translate into a diversity gain. We then introduce two efficient turbo
packet combining algorithms that are inspired by minimum mean square
error (MMSE)-based turbo equalization techniques. Both schemes can
be viewed as low-complexity versions of the optimal MAP turbo combiner.
The first scheme is called \emph{signal-level} turbo combining and
performs packet combining and multiple transmission ISI cancellation
jointly at the signal-level. The second scheme, called \emph{symbol-level}
turbo combining, allows ARQ rounds to be separately turbo equalized,
while combining is performed at the filter output. We conduct a complexity
analysis where we demonstrate that both algorithms have almost the
same computational cost as the conventional log-likelihood ratio (LLR)-level
combiner. Simulation results show that both proposed techniques outperform
LLR-level combining, while for some representative MIMO configurations,
signal-level combining has better ISI cancellation capability and
achievable diversity order than that of symbol-level combining.\end{abstract}

\section{Introduction\label{sec:Introduction}}

\subsection{Research Motivation}

\PARstart{H}{ybrid--automatic} repeat request (ARQ) protocols and
multiple-input--multiple-output (MIMO) play a key role in the evolution
of current wireless systems toward high data rate wireless broadband
standards \cite{PeisaEricsson_VTCS07}. While MIMO techniques allow
the space and time diversities of the multi-antenna channel to be
translated into diversity and/or multiplexing gains \cite{Woliansky},
hybrid--ARQ mechanisms exploit the ARQ delay, i.e., the maximum number
of ARQ transmission rounds, to reduce the frame error rate (FER) and
therefore increase the system throughput \cite{ChaseTComm1985,Harvey_WickerTComm1994}. 

In the last few years, special interest has been paid to the joint
design of the transmission combiner (also referred to as \emph{{}``packet
combiner''}) and the signal processor (detection and/or equalization)
receiver. Combining schemes targeting a joint design approach were
first proposed by Samra and Ding for single antenna systems operating
over intersymbol interference (ISI) channels \cite{SamraDing_ICASSP_02,SamraDing_ISIT02,SamraDing_Asilomar2002,Samra_DingIEqual_TComm05},
and are called transmission combining with integrated equalization
(IEQ). In particular, it was shown in \cite{Samra_DingIEqual_TComm05}
that, when concatenated with an outer code, IEQ performs better than
the iterative combining scheme introduced by Doan and Narayanan \cite{Doan_Narayanan_ItertCombTComm02}.
In iterative combining, multiple copies of the same packet are independently
interleaved and combining is performed by iterating between multiple
equalizers before channel decoding. The IEQ concept was then extended
to MIMO systems with flat fading to jointly perform co-antenna interference
(CAI) cancellation and transmission combining \cite{Dabak_ICC03,Samra_Ding04,Ding_MIMOARQ_Sphere_SigProc}.
In parallel, several other MIMO ARQ architectures exploiting the high
degree of freedom in the design of the MIMO ARQ transmitter were proposed
(e.g. \cite{ZhizhongDing_RiceICC03,Koike_ICC04,H_Zheng_et_al_PIMRC02,Hottinen_et_al_ISSPA03,Ibi_el_al_CommLett_06,Krishnaswamy_VTCfall06,Jang_et_al_ISIT07,Ibi_el_al_TVT_07}).
Turbo coded ARQ schemes with iterative minimum mean square error (MMSE)
frequency domain equalization (FDE) for single carrier transmission
over broadband channel were proposed for direct sequence code division
multiple access (DS-CDMA) and MIMO systems in \cite{Garg_Adachi_JSAC06}
and \cite{AdachiVTC2005,Adachi_HARQ_SCMIMO_VTC06}, respectively. 

Recently, in a seminal paper by El Gamal \emph{et al}. \cite{DMD_Gamal_IT_2006},
the diversity--multiplexing tradeoff %
\footnote{A fundamental tool for the design of space--time coding/multiplexing
architectures initially proposed by Zheng and Tse for flat fading
\cite{Zheng-Tse}, and later extended to frequency selective fading
\cite{Medles-Slock_ISIT05,Slock_ITW2007,Bolskei_ISIT07}.%
} of the MIMO ARQ flat fading channel was characterized, and was referred
to as diversity--multiplexing--delay tradeoff. The authors proved
that the ARQ delay presents an important source of diversity even
when the channel is constant over ARQ transmission rounds, a scenario
referred to as long-term static channel. In particular, it was shown
that operating over such a channel with a large ARQ delay results
in a flat diversity--multiplexing tradeoff. This means that one can
achieve full diversity and multiplexing gains if large ARQ windows
are allowed. The diversity--multiplexing--delay tradeoff was then
investigated in the case of delay-sensitive services and block-fading
MIMO channels in \cite{Holliday_Glodsmith_Poor_ICC06} and \cite{Chuang_et_al_DMD_IT08},
respectively.

\subsection{In this Paper}

Motivated by the IEQ concept \cite{Samra_DingIEqual_TComm05} and
the results in \cite{DMD_Gamal_IT_2006}, we investigate efficient
IEQ-aided packet combining strategies for coded transmission with
hybrid--ARQ operating over MIMO-ISI channels. Our main objective is
to reduce the number of ARQ rounds required to correctly decode a
data packet while keeping the receiver complexity (computational load
and memory requirements) affordable. In our design, packet combining
is performed at each ARQ round by exchanging soft information in an
iterative (turbo) fashion between the \emph{soft packet combiner}
and the soft-input--soft-output (SISO) decoder. We refer to this combining
family as \emph{{}``turbo packet combining''.} 

We focus on space--time bit-interleaved coded modulation (ST-BICM)
transmitter schemes with Chase-type ARQ, i.e., the data packet is
entirely retransmitted. The choice of ST-BICM is motivated by the
simplicity of this coding scheme, and the efficiency of its iterative
decoding (ID) receiver in achieving high diversity and coding gains
over block-fading MIMO-ISI channels \cite{Ariyavisitakul_ICC00,Tonello,Visoz_TComm03,vandendorpe_SigProc_04,Visoz-groupMMSE-journal,Ait-idir_TVT}.
Our work is still valid for other space--time codes (STCs). Note that
some practical systems employ hybrid--ARQ with incremental redundancy
(IR). In IR-type ARQ, retransmissions only carry portions of the data
packet. It presents an efficient technique for increasing the system
throughput while keeping the error performance acceptable. In this
paper, we restrict our work to Chase-type ARQ. Turbo combining techniques
for broadband MIMO transmission with IR-type ARQ are left for future
investigations. 

First of all, we derive the optimal \emph{maximum a posteriori} (MAP)
turbo packet combining algorithm %
\footnote{In this paper, optimality refers to the exploitation of delay, space,
time, and multipath diversities of the MIMO-ISI ARQ channel to combine
multiple transmissions. %
} that makes use of all diversities available in the MIMO-ISI ARQ channel
to perform transmission combining. The turbo packet combining strategies
we introduce in this paper can be seen as low-complexity sub-optimal
techniques of the MAP combining algorithm. An important ingredient
in MAP turbo combining is an analogy between multiple transmissions
and antennas, and which consists of considering ARQ rounds as virtual
receive antennas. This allows the ARQ delay, i.e., maximum number
of ARQ rounds, to be translated into receive diversity. We then analyze
the outage performance of the MIMO-ISI ARQ channel. This analysis
allows us to know how the ARQ delay influences the outage probability
of the MIMO ARQ system. It also serves as a theoretical foundation
for the turbo packet combiners we propose in this paper. We also investigate
the outage-based power loss due to multiple transmission rounds. This
analysis establishes that in the outage region of interest (corresponding
to an outage between $10^{-2}$ and $10^{-3}$) the power loss due
to ARQ is below $0.25$dB. 

The next step in our work corresponds to the derivation of two turbo
packet combining strategies for the MIMO-ISI ARQ channel. Both techniques
are inspired by the unconditional MMSE turbo equalization schemes
of \cite{vandendorpe_SigProc_04} and \cite{Tuchler_TSP_2002}. The
first algorithm, named \emph{signal-level} turbo packet combining,
presents a low-complexity version of MAP turbo combining. It performs
packet combining and equalization using signals from all transmission
rounds. In contrast to what was initially stated in \cite{Ait-Idir_WCNC_08},
we show that the computational complexity of this scheme is less sensitive
to the number of ARQ rounds. Moreover, we provide an optimized implementation
where it is not necessary for the receiver to store all signal vectors
and channel matrices. The second combining scheme, namely, \emph{symbol-level}
turbo combining, performs soft equalization separately for each round,
and combines multiple transmissions at the level of filter outputs.
It has the same computational complexity and fewer memory requirements
compared with the first scheme. We also show that receiver requirements
(computational complexity and memory) of both turbo combining schemes
are almost similar to those of conventional log-likelihood ratio (LLR)-level
combining, where extrinsic LLRs corresponding to multiple transmissions
are simply added together before SISO decoding. Finally, we provide
numerical simulations for some MIMO configurations demonstrating the
superior performance of the proposed algorithms compared with LLR-level
combining, and the significant gains they offer with respect to both
the outage probability and the matched filter bound (MFB). 

Throughout the paper, the following notation is used. Superscript
$^{\top}$ denotes transpose, and $^{H}$ denotes Hermitian transpose.
$\mathbb{{E}}\left[.\right]$ is the mathematical expectation of the
argument $\left(.\right)$. When $\mathbf{X}$ is a square matrix,
$\mathrm{det}\left(\mathbf{X}\right)$ denotes the determinant of
$\mathbf{X}$. For each complex vector $\mathbf{x}\in\mathbb{C}^{N}$,
$\mathrm{diag}\left\{ \mathbf{x}\right\} $ is the $N\times N$ diagonal
matrix whose diagonal entries are the elements of $\mathbf{x}$. $\mathbf{I}_{N}$
is the $N\times N$ identity matrix, and $\mathbf{0}_{N\times Q}$
denotes an all zero $N\times Q$ matrix. $\otimes$ is the Kronecker
product, and $j=\sqrt{-1}$. 

The following sections of the paper are organized as follows. In Section
\ref{sec:HARQ-Comm-Model}, we provide a description of the MIMO ARQ
system model and introduce some assumptions considered in this paper.
In Section \ref{sec:OptimalPacketComb_and_Outage}, we derive the
structure of the optimal MAP turbo combining scheme, and analyze the
outage probability and the outage-based power loss of the considered
MIMO ARQ system. Section \ref{sec:Proposed_Combiners} details the
structure of the proposed combining schemes and discusses complexity
issues. Numerical results are provided in Section \ref{sec:Numerical-Results}.
The paper is concluded in Section \ref{sec:Conclusions}.

\section{System Model and Assumptions \label{sec:HARQ-Comm-Model}}

\begin{figure*}[t]
\noindent \begin{centering}
\includegraphics[scale=0.6]{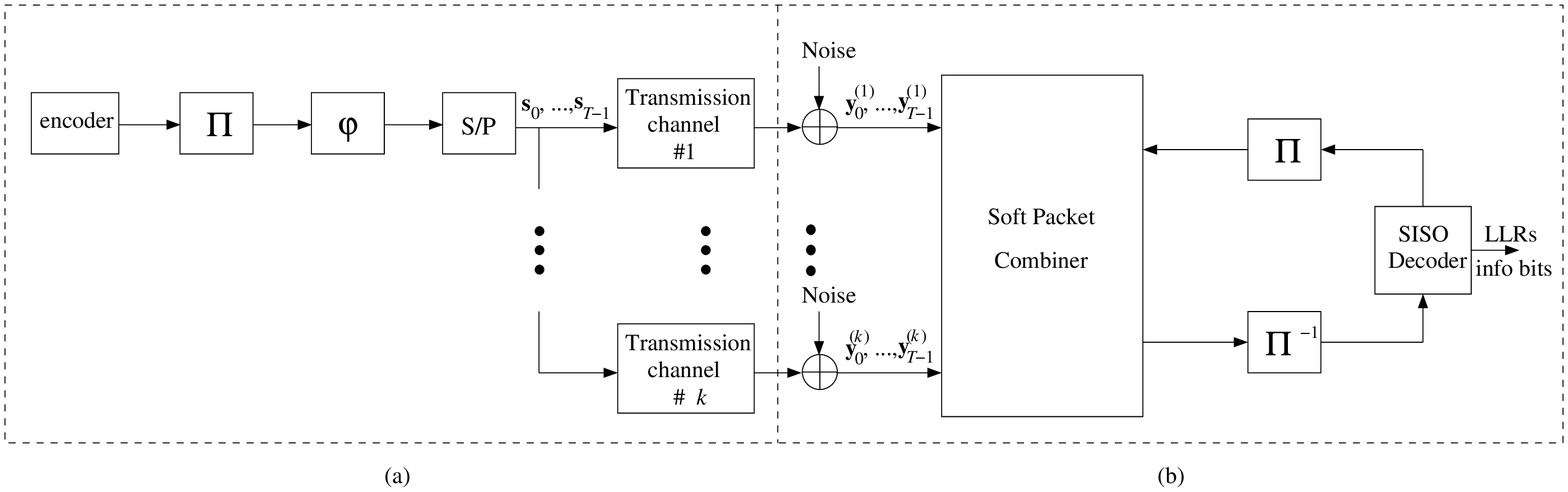}
\par\end{centering}

\caption{\label{fig:STBICM_ARQ} ST-BICM diagram with ARQ and turbo packet
combining: (a) transmitter, (b) receiver.}

\end{figure*}

We consider a multi-antenna link operating over a frequency selective
fading channel and using an ARQ protocol at the upper layer. The transmitter
and the receiver are equipped with $N_{T}$ transmit and $N_{R}$
receive antennas, respectively. The MIMO-ISI channel is composed of
$L$ taps (index $l=0,\cdots,L-1$). Each data stream is encoded with
the aid of a $\rho$-rate channel encoder, interleaved using a semi-random
interleaver $\Pi$, then modulated and space--time multiplexed over
the $N_{T}$ transmit antennas. This presents a ST-BICM coding scheme.
The mapping function that relates each set of $M$ coded and interleaved
bits $b_{1,t,i},\cdots,b_{M,t,i}$ to a symbol $s_{t,i}$ that belongs
to the constellation set $\mathcal{S}$ is denoted $\varphi:\,\left\{ 0,1\right\} ^{M}\rightarrow\mathcal{S}$,
where $t=1,\cdots,N_{T}$, and $i=0,\cdots,T-1$ are the transmit
antenna and the channel use indices, respectively, and $M=\log_{2}\left|\mathcal{S}\right|$.
The $N_{T}\times T$ symbol matrix corresponding to the entire frame
is denoted\begin{equation}
\mathbf{S}\triangleq\left[\mathbf{s}_{0},\cdots,\mathbf{s}_{T-1}\right]\in\mathcal{S}^{N_{T}\times T},\label{eq:Symbol_Mat_Def}\end{equation}
\begin{equation}
\mathbf{s}_{i}\triangleq\left[\mathbf{s}_{1,i},\cdots,s_{N_{T},i}\right]^{\top}\in\mathcal{S}^{N_{T}}\label{eq:Symbol_Vec_Def}\end{equation}
is the vector of transmitted symbols at time instant $i$. The rate
of this transmission scheme is therefore $R=\rho MN_{T}$. When the
transmitter receives a negative acknowledgment (NACK) message due
to an erroneously decoded block, subsequent transmission rounds occur
until the packet is correctly received or a preset maximum number
of rounds, i.e., ARQ delay, $K$ is reached. The round index is denoted
$k=1,\cdots,K$. Reception of a positive acknowledgment (ACK) indicates
a successful decoding and the transmitter moves on to the next block
message. We suppose that the signaling channel carrying the one bit
ACK/NACK feedback message is error free. In addition, we assume perfect
packet error detection (typically, using a cyclic redundancy check
(CRC) code). Therefore, a decoding failure corresponds to an erroneous
decoding outcome after $K$ rounds. We focus on Chase-type ARQ mechanisms,
i.e., the symbol matrix $\mathbf{S}$ is completely retransmitted.
Both puncturing and mapping diversity, i.e., optimization of the mapping
function over transmission rounds, are not investigated in this paper,
and are left for future contributions. We use a zero padding (ZP)
sequence $\mathbf{0}_{N_{T}\times L}$ to prevent inter-block interference
(IBI). The ST-BICM scheme with ARQ is depicted in Fig. \ref{fig:STBICM_ARQ}.
a. The MIMO-ISI channel is assumed to be quasi-static block fading,
i.e., constant over a frame that spans $T$ channel use and independently
changes from round to round. This scenario corresponds to the so-called
short-term static channel case where ARQ transmission rounds see different
and independent channel realizations \cite{DMD_Gamal_IT_2006}. The
long-term static channel corresponds to the case where the channel
is constant over all rounds related to the transmission of the same
information block, i.e., $\mathbf{H}_{l}^{\left(k\right)}=\mathbf{H}_{l}\,\,\forall k\in\left\{ 1,\cdots,K\right\} $.
Note that in orthogonal frequency division multiplexing (OFDM) broadband
wireless systems, the ARQ channel is rather short-term static because
frequency hopping is used to mitigate ISI. While in time division
multiplexing (TDM)-based systems, the channel dynamic can be either
short or long-term static depending on the Doppler spread. In addition,
we suppose that the channel profile, i.e., number of paths and power
distribution, is identical for at least $K$ consecutive rounds. This
is a reasonable assumption for slowly time-varying wireless fading
channels because the channel profile dynamic is mainly related to
the shadowing effect. At the $k$th round, the channel impulse response
is represented by the $N_{R}\times N_{T}$ complex matrices $\mathbf{H}_{0}^{\left(k\right)},\cdots,\mathbf{H}_{L-1}^{\left(k\right)}$
corresponding respectively to taps $0,\ldots,L-1$, and whose entries
are zero-mean circularly symmetric Gaussian $h_{r,t,l}^{\left(k\right)}\sim\mathcal{CN}\left(0,\sigma_{l}^{2}\right)$,
where $h_{r,t,l}^{\left(k\right)}$ denotes the $\left(r,t\right)$th
element of matrix $\mathbf{H}_{l}^{\left(k\right)}$. The total energy
of taps $l=0,\cdots,L-1$ is normalized to one, i.e., $\sum_{l=0}^{L-1}\sigma_{l}^{2}=1.$
Therefore, the channel energy per receive antenna $r=1,\cdots,N_{R}$
is%
 \begin{equation}
{\displaystyle \sum_{l=0}^{L-1}}{\displaystyle \sum_{t=1}^{N_{T}}}\,\mathbb{{E}}\left[\left|h_{r,t,l}^{\left(k\right)}\right|^{2}\right]=N_{T}.\label{eq:CH_EnergyNormlize}\end{equation}
We suppose that no channel knowledge is available at the transmitter.
Equal power transmission turns out to be the best power allocation
strategy. In addition, under the assumption of infinitely deep interleaving,
and by normalizing the symbol energy to one, we get \begin{equation}
\mathbb{{E}}\left[\mathbf{s}_{i}\mathbf{s}_{i}^{H}\right]=\mathrm{\mathbf{I}}_{N_{T}}.\label{eq:TxPowerNormalize}\end{equation}

\noindent \begin{flushleft}
At the $k$th round, after down-conversion and sampling at the symbol
rate, the baseband complex received signal on the $r$th antenna and
at time instant $i$ is\begin{equation}
y_{r,i}^{\left(k\right)}=\sum_{l=0}^{L-1}\sum_{t=1}^{N_{T}}h_{r,t,l}^{\left(k\right)}s_{t,i-l}+n_{r,i}^{\left(k\right)},\label{eq:SignalModel_kthTx}\end{equation}
 where $n_{r,i}^{\left(k\right)}$ is the noise on the $r$th antenna,
and $\mathbf{n}_{i}^{\left(k\right)}\triangleq\left[n_{1,i}^{\left(k\right)},\cdots,n_{N_{R},i}^{\left(k\right)}\right]^{\top}\sim\mathcal{CN}\left(\mathbf{0}_{N_{R}\times1},\sigma^{2}\mathrm{\mathbf{I}}_{N_{R}}\right)$.
\par\end{flushleft}

\section{Optimal Turbo Packet Combining and Outage Analysis\label{sec:OptimalPacketComb_and_Outage}}

In this section, we provide a brief description of the structure of
the turbo packet combining concept we propose in this paper, and introduce
the optimal MAP turbo combiner. We also investigate the outage probability
and the outage-based transmit power loss then provide a numerical
analysis.

\subsection{General Architecture and Optimal Turbo Combining\label{sub:GeneralArch_MAPcomb}}

The turbo packet combining strategies we propose in this paper allow
decoding of a data packet transmitted over multiple MIMO-ISI channels
in an iterative (turbo) fashion through the exchange of extrinsic
information between the soft packet combiner and the SISO decoder.
The main difference with conventional LLR-based packet combining is
that multiple transmissions are combined before the computation of
the soft information using a SISO packet combiner, while in LLR-level
combining the soft outputs of different ARQ rounds are simply added
together before channel decoding. The general block diagram is depicted
in Fig. \ref{fig:STBICM_ARQ}. b. Let $N$ denote the number of turbo
iterations performed between the combiner and the decoder at the $k$th
round (index $n=1,\cdots,N$), and
\begin{figure*}[!t]
\normalsize 
\setcounter{mytempeqncnt}{\value{equation}} 
\setcounter{equation}{7}
\begin{equation} \phi_{m,t,i,n}^{e}=\log\frac{\mathrm{Pr}\left\{ \mathbf{y}^{\left(k\right)}\mid b_{m,t,i}=1\,;\,\mathbf{H}_{0}^{\left(1\right)},\cdots,\mathbf{H}_{L-1}^{\left(k\right)},\, a\, priori\,\,\mathrm{LLRs}\right\} }{\mathrm{Pr}\left\{ \mathbf{y}^{\left(k\right)}\mid b_{m,t,i}=0\,;\,\mathbf{H}_{0}^{\left(1\right)},\cdots,\mathbf{H}_{L-1}^{\left(k\right)},\, a\, priori\,\,\mathrm{LLRs}\right\} },\label{eq:EXT_LLR_MAP}\end{equation}
\setcounter{equation}{\value{mytempeqncnt}} 
\hrulefill 
\vspace*{4pt} \end{figure*}\begin{figure*}[!t]
\normalsize 
\setcounter{mytempeqncnt}{\value{equation}} 
\setcounter{equation}{13}
\begin{equation} \phi_{m,t,i,n}^{e}=\log\frac{{\displaystyle \sum_{\mathbf{s}\in\mathcal{S}_{m,t,i}^{1}}}\exp\left\{ -\frac{1}{2\sigma^{2}}\left\Vert \mathbf{y}^{\left(k\right)}-\mathbf{H}^{\left(k\right)}\mathbf{s}\right\Vert ^{2}+{\displaystyle \sum_{\left(m',t',i'\right)\neq\left(m,t,i\right)}}\varphi_{m'}^{-1}\left(x_{t',i'}\right)\phi_{m',t',i',n}^{a}\right\} }{{\displaystyle \sum_{\mathbf{s}\in\mathcal{S}_{m,t,i}^{0}}}\exp\left\{ -\frac{1}{2\sigma^{2}}\left\Vert \mathbf{y}^{\left(k\right)}-\mathbf{H}^{\left(k\right)}\mathbf{s}\right\Vert ^{2}+{\displaystyle \sum_{\left(m',t',i'\right)\neq\left(m,t,i\right)}}\varphi_{m'}^{-1}\left(x_{t',i'}\right)\phi_{m',t',i',n}^{a}\right\} },\label{eq:EXT_LLR_MAP_expr2}\end{equation}
\setcounter{equation}{5} 
\hrulefill 
\vspace*{4pt} \end{figure*}

\begin{align}
\,\,\,\,\,\,\,\,\boldsymbol{{\phi}}_{t,i,n}^{e}\triangleq & \left[\phi_{1,t,i,n}^{e},\cdots,\phi_{M,t,i,n}^{e}\right]^{\top}\in\mathbb{R}^{M},\nonumber \\
 & \,\,\,\,\,\,\,\,\,\,\,\,\,\,\,\,\,\,\,\,\,\,\,\,\,\left(t,i\right)\in\left\{ 1,\cdots,N_{T}\right\} \times\left\{ 0,\cdots,T-1\right\} \label{eq:Ext_Vec_Def}\end{align}

\noindent denote the vectors of extrinsic log-likelihood ratio (LLR)
values generated by the soft combiner at iteration $n$. $\phi_{m,t,i,n}^{e}$
is the extrinsic information related to coded and interleaved bit
$b_{m,t,i}$ at turbo iteration $n$. We similarly define \emph{a
priori} vectors\[
\boldsymbol{{\phi}}_{t,i,n}^{a}\triangleq\left[\phi_{1,t,i,n}^{a},\cdots,\phi_{M,t,i,n}^{a}\right]^{\top}\in\mathbb{R}^{M},\]
 available at the input of the soft combiner at iteration $n$. For
the sake of notation simplicity, the round index is not used in LLRs.
At the $n$th iteration of the $k$th round, the soft packet combiner
makes use of the $N_{T}T$ \emph{a priori} vectors $\boldsymbol{{\phi}}_{1,0,n}^{a},\cdots,\boldsymbol{{\phi}}_{N_{T},T-1,n}^{a}$
and received signals to combine transmissions corresponding to rounds
$1,\cdots,k$, and compute extrinsic vectors $\boldsymbol{{\phi}}_{1,0,n}^{e},\cdots,\boldsymbol{{\phi}}_{N_{T},T-1,n}^{e}$.
These extrinsic LLRs are de-interleaved and sent to the SISO decoder
to compute \emph{a posteriori} information about useful bits and extrinsic
LLRs about coded bits. The generated extrinsic information is then
interleaved and fed back to the soft combiner to serve as \emph{a
priori} information $\boldsymbol{{\phi}}_{1,0,n+1}^{a},\cdots,\boldsymbol{{\phi}}_{N_{T},T-1,n+1}^{a}$
at next iteration $n+1$. Note that the feedback of a NACK message
does not necessarily mean that all information bits are erroneous.
Therefore, extrinsic information generated by the SISO decoder during
the last iteration of ARQ round $k-1$ can be used as \emph{a priori
}information at the first iteration of ARQ round $k$. %
\footnote{Generally speaking, iterative processing at round $k$ will help correct
information bits erroneously decoded during round $k-1$, while the
LLR values of other bits remain the same.%
}

Now, let us focus on the optimal soft packet combiner that allows
the exploitation of all diversities, i.e., space, time, multipath,
and retransmission, present in the MIMO-ISI ARQ channel to iteratively
compute extrinsic information about coded and interleaved bits. First,
let us introduce\begin{equation}
\mathbf{y}_{i}^{\left(k\right)}\triangleq\left[y_{1,i}^{\left(k\right)}\cdots y_{N_{R},i}^{\left(k\right)}\right]^{\top}\label{eq:rx_signalVec}\end{equation}
that groups the signals received at time instant $i$ of the $k$th
round (\ref{eq:SignalModel_kthTx}). We assume that the signals received
at rounds $1,\cdots,k$ (i.e., $\mathbf{y}_{0}^{\left(1\right)},\cdots,\mathbf{y}_{T-1}^{\left(k\right)}$)
and their corresponding channel responses (i.e., $\mathbf{H}_{0}^{\left(1\right)},\cdots,\mathbf{H}_{L-1}^{\left(k\right)}$)
are available at the receiver. Note that this assumption may present
an important limiting factor (in addition to the computational complexity)
for implementing the optimal turbo combiner, since all signals and
channel responses have to be stored in the receiver. The low-complexity
signal-level turbo combining strategy we introduce in Section \ref{sec:Proposed_Combiners}
relaxes this condition by using two recursions for keeping signals
and channel matrices of previous rounds. At the $n$th iteration of
round $k$, the optimal soft combiner computes extrinsic LLR about
coded and interleaved bit $b_{m,t,i}$ according to the MAP criterion
(\ref{eq:EXT_LLR_MAP}), where\setcounter{equation}{8}\begin{equation}
\mathbf{y}^{\left(k\right)}\triangleq\left[\mathbf{y}_{T-1}^{\left(1\right)^{\top}},\cdots,\mathbf{y}_{T-1}^{\left(k\right)^{\top}},\cdots,\mathbf{y}_{0}^{\left(1\right)^{\top}},\cdots,\mathbf{y}_{0}^{\left(k\right)^{\top}}\right]^{\top}\in\mathbb{C}^{kN_{R}T}.\label{eq:Comb_RxSig_MAP}\end{equation}
Note that this vector representation is of a great importance because
it allows us to view each transmission round as a source of an additional
set of virtual $N_{R}$ receive antennas. Therefore, ARQ diversity
translates into space diversity (i.e., virtual receive antennas).
The signal vector $\mathbf{y}^{\left(k\right)}$ corresponding to
the transmission of matrix $\mathbf{S}$ over $k$ MIMO-ISI channels
can be expressed as,

\begin{equation}
\mathbf{y}^{\left(k\right)}=\mathbf{H}^{\left(k\right)}\mathbf{s}+\mathbf{n}^{\left(k\right)},\label{eq:CommModel_OptimalMAP}\end{equation}
where $\mathbf{H}^{\left(k\right)}$ is a $kN_{R}T\times N_{T}T$
block Toeplitz matrix, 

\begin{equation}
\mathbf{H}^{\left(k\right)}\triangleq\left[\begin{array}{ccccc}
\begin{array}{|c|}
\hline \,\mathbf{H}_{0}^{\left(1\right)}\,\,\\
\vdots\\
\,\mathbf{H}_{0}^{\left(k\right)}\,\,\\\hline \end{array} & \cdots & \begin{array}{|c|}
\hline \mathbf{H}_{L-1}^{\left(1\right)}\\
\vdots\\
\mathbf{H}_{L-1}^{\left(k\right)}\\\hline \end{array}\\
 & \ddots &  & \ddots\\
 &  & \begin{array}{|c|}
\hline \,\mathbf{H}_{0}^{\left(1\right)}\,\,\\
\vdots\\
\,\mathbf{H}_{0}^{\left(k\right)}\,\,\\\hline \end{array} & \cdots & \begin{array}{|c|}
\hline \mathbf{H}_{L-1}^{\left(1\right)}\\
\vdots\\
\mathbf{H}_{L-1}^{\left(k\right)}\\\hline \end{array}\end{array}\right],\label{eq:BlockMat_MAP}\end{equation}
and\begin{equation}
\mathbf{s}\triangleq\left[\mathbf{s}_{T-1}^{\top},\cdots,\mathbf{s}_{0}^{\top}\right]^{\top}\in\mathcal{S}^{N_{T}T},\label{eq:SymbolVec_MAP}\end{equation}
\begin{equation}
\mathbf{n}^{\left(k\right)}\triangleq\left[\mathbf{n}_{T-1}^{\left(1\right)^{\top}},\cdots,\mathbf{n}_{T-1}^{\left(k\right)^{\top}},\cdots,\mathbf{n}_{0}^{\left(1\right)^{\top}},\cdots,\mathbf{n}_{0}^{\left(k\right)^{\top}}\right]^{\top}\in\mathbb{C}^{kN_{R}T}.\label{eq:NoiseVec_MAP}\end{equation}
With respect to (\ref{eq:CommModel_OptimalMAP}), extrinsic LLR given
by (\ref{eq:EXT_LLR_MAP}) can now be expressed according to (\ref{eq:EXT_LLR_MAP_expr2}),
where $\mathcal{S}_{m,t,i}^{b}=\left\{ \mathbf{s}\in\mathcal{S}^{N_{T}T}\mid\varphi_{m}^{-1}\left(s_{t,i}\right)=b\right\} ,\,\, b=0,\,1$.

\subsection{Outage Probability and Outage-Based Transmit Power Loss}

It is well known that for non-ergodic channels, i.e., block fading
quasi-static channels, outage-probability $P_{out}$ \cite{Shamai_Ozarow_WynerIT94,Telatar_EurpTelecom_99,Foshini_Gans_WPC_98}
is regarded as a meaningful tool for performance evaluation because
it provides a lower bound on the block error rate (BLER) \cite[p. 187]{Tse_Viswanath_Book}.
The outage probability is defined as the probability that the mutual
information, as a function of the channel realization and the average
signal to noise ratio (SNR) $\gamma$ per receive antenna, is below
the transmission rate $R$. Mutual information rates of quasi-static
frequency selective fading MIMO channel have been investigated in
\cite{ElGamal_IT04,Duman_Tcomm04}.

\subsubsection{Outage Probability }

To derive the outage probability of the considered MIMO ARQ system,
we use the \emph{renewal theory} \cite{RenewalTheory_Wolff} which
was first used by Zorzi and Rao to analyze the performance of ARQ
protocols \cite{RenewalTheory_Zorzi}. Recently, it was also used
by \cite{Caire_IT01,DMD_Gamal_IT_2006} to evaluate the performance
of ARQ systems operating over wireless flat fading channels. Let $\mathcal{A}_{k}$
denote the event that an ACK message is fed back at round $k$, and
$\mathcal{E}_{k}$ the event that the ARQ system is in outage at round
$k$. Under the assumption of perfect packet error detection and error-free
ACK/NACK feedback, and by applying the \emph{renewal theory}, the
outage probability for a given SNR $\gamma$ and target rate $R$
is given as\setcounter{equation}{14}\begin{align}
P_{out}^{R}\left(\gamma\right) & =\Pr\left\{ \mathcal{E}_{K},\bar{\mathcal{A}}_{1},\cdots,\bar{\mathcal{A}}_{K-1}\right\} .\label{eq:Outage_RenewalTheory}\end{align}
Note that a Chase-type ARQ mechanism with an ARQ delay $K$ can be
viewed as a repetition coding scheme where $K$ parallel sub-channels
are used to transmit one symbol message \cite[p. 194]{Tse_Viswanath_Book}.
Therefore, (\ref{eq:Outage_RenewalTheory}) can be expressed as\begin{multline} P_{out}^{R}\left(\gamma\right)=\Pr\left\{ \frac{1}{K}I\left(\mathbf{s};\mathbf{y}^{\left(K\right)}\mid\mathbf{H}^{\left(K\right)},\gamma\right)<R,\right.\,\,\,\,\\ \bar{\mathcal{A}}_{1},\cdots,\bar{\mathcal{A}}_{K-1}\biggr\}.\label{eq:Outage_eq2}\end{multline}
The virtual $KN_{R}\times N_{T}$ MIMO-ISI communication model at
the $K$th ARQ round is\[
\left[\begin{array}{c}
\mathbf{y}_{i}^{\left(1\right)}\\
\vdots\\
\mathbf{y}_{i}^{\left(K\right)}\end{array}\right]=\sum_{l=0}^{L-1}\left[\begin{array}{c}
\mathbf{H}_{l}^{\left(1\right)}\\
\vdots\\
\mathbf{H}_{l}^{\left(K\right)}\end{array}\right]\mathbf{s}_{i-l}+\left[\begin{array}{c}
\mathbf{n}_{i}^{\left(1\right)}\\
\vdots\\
\mathbf{n}_{i}^{\left(K\right)}\end{array}\right],\]
and the mutual information $I\left(\mathbf{s};\mathbf{y}^{\left(K\right)}\mid\mathbf{H}^{\left(K\right)},\gamma\right)$
in (\ref{eq:Outage_eq2}) can therefore be expressed in the case of
i.i.d circularly symmetric complex Gaussian channel inputs as in \cite{ElGamal_IT04},
i.e.,\begin{multline}
I\left(\mathbf{s};\mathbf{y}^{\left(K\right)}\mid\mathbf{H}^{\left(K\right)},\gamma\right)=\,\,\,\,\,\,\,\,\,\,\,\,\,\,\,\,\,\,\,\,\,\,\,\,\,\,\,\,\,\,\,\,\,\,\,\,\,\,\,\,\,\,\,\\
\frac{1}{T}\sum_{i=0}^{T-1}\log_{2}\left(\det\left(\mathbf{I}_{KN_{R}}+\frac{\gamma}{N_{T}}\boldsymbol{{\Lambda}}_{i}^{\left(K\right)}\boldsymbol{{\Lambda}}_{i}^{\left(K\right)^{H}}\right)\right),\label{eq:MutualInfo_ExactGauss}\end{multline}
where $\boldsymbol{{\Lambda}}_{i}^{\left(K\right)}$ is the discrete
Fourrier transform (DFT) of the $K$th round $KN_{R}\times N_{T}$
virtual MIMO-ISI channel at the $i$th frequency bin, i.e.,

\begin{equation}
\boldsymbol{{\Lambda}}_{i}^{\left(K\right)}=\sum_{l=0}^{L-1}\left[\begin{array}{c}
\mathbf{H}_{l}^{\left(1\right)}\\
\vdots\\
\mathbf{H}_{l}^{\left(K\right)}\end{array}\right]\exp\left\{ -j\frac{2\pi}{T}il\right\} .\label{eq:DFT_VirtualChannel}\end{equation}

\subsubsection{Outage-Based Transmit Power Loss}

To compare the outage probability performance of different ARQ configurations
that operate at the same rate $R$ but use different ARQ delays, we
consider a short-term power constraint scenario where the same power
$\Gamma$ is used for all transmission rounds, i.e., the $k$th round
transmit power is $\Gamma_{k}=\Gamma\,\,\forall k$. We evaluate the
power loss incurred by multiple transmission rounds due to link outage.
Note that system performance can be improved when a power control
algorithm is jointly used with packet combining (typically, a long-term
power constraint scenario), but this is beyond the scope of this paper.
The average SNR present in the outage expression (\ref{eq:Outage_eq2})
is therefore given as\begin{equation}
\gamma=\Gamma\frac{N_{T}}{\sigma^{2}}.\label{eq:SNR_expr}\end{equation}
Let $p$ count the number of information blocks, $q=1,\cdots,p$ denote
the block index, and $\mathcal{T}_{q}$ the number of rounds used
for transmitting block $q$. Therefore, for a given ARQ delay $K$,
average SNR $\gamma$, and rate $R$, the average transmit power is\begin{align}
\Gamma_{avg} & =\lim_{p\rightarrow\infty}\frac{\sum_{q=1}^{p}\mathcal{T}_{q}}{p}\Gamma\nonumber \\
 & =\mathbb{{E}}\left[\mathcal{T}\mid K,\gamma,R\right]\Gamma.\label{eq:AverageTxPower}\end{align}
This indicates that an ARQ protocol with an ARQ delay $K$ and operating
with rate $R$ at average SNR $\gamma$ incurs an \emph{outage-based
transmit power loss} of $10\log_{10}\left(\mathbb{{E}}\left[\mathcal{T}\mid K,\gamma,R\right]\right)$
compared with an ARQ with $K=1$ round (i.e., no retransmissions).

\subsection{Outage Analysis\label{sub:Outage-Analysis}}

In the following subsection we investigate, using simulations, both
the outage probability and the outage-based transmit power loss for
some MIMO-ISI ARQ configurations. This will serve as a theoretical
foundation for the performance evaluation of turbo packet combiners
which we will introduce in the next subsection. Let us consider a
MIMO-ISI channel with $L=2$ taps and equally distributed power, i.e.,
$\sigma_{0}^{2}=\sigma_{1}^{2}=\frac{1}{2}$. We use Monte Carlo simulations
to evaluate the outage probability (\ref{eq:Outage_eq2}) of the considered
ARQ system. We choose $T=256$ channel use. At each round $k$, a
$N_{R}\times N_{T}$ MIMO-ISI channel $\mathbf{H}_{0}^{\left(k\right)}$
and $\mathbf{H}_{1}^{\left(k\right)}$ is generated, and the mutual
achievable rate after $k$ rounds is computed using (\ref{eq:MutualInfo_ExactGauss}).
If the target rate $R$ is not reached and $k<K$, the system moves
on to the next round $k+1$. The ARQ process is stopped and another
is started, either because of system outage (i.e., the achievable
rate after $K$ rounds is below $R$) or non-outage (i.e., the achievable
rate is greater than $R$ after round $k\leq K$). %
\begin{figure*}[t]
\begin{raggedright}
\hfill{}\includegraphics[width=7.5cm,height=9cm]{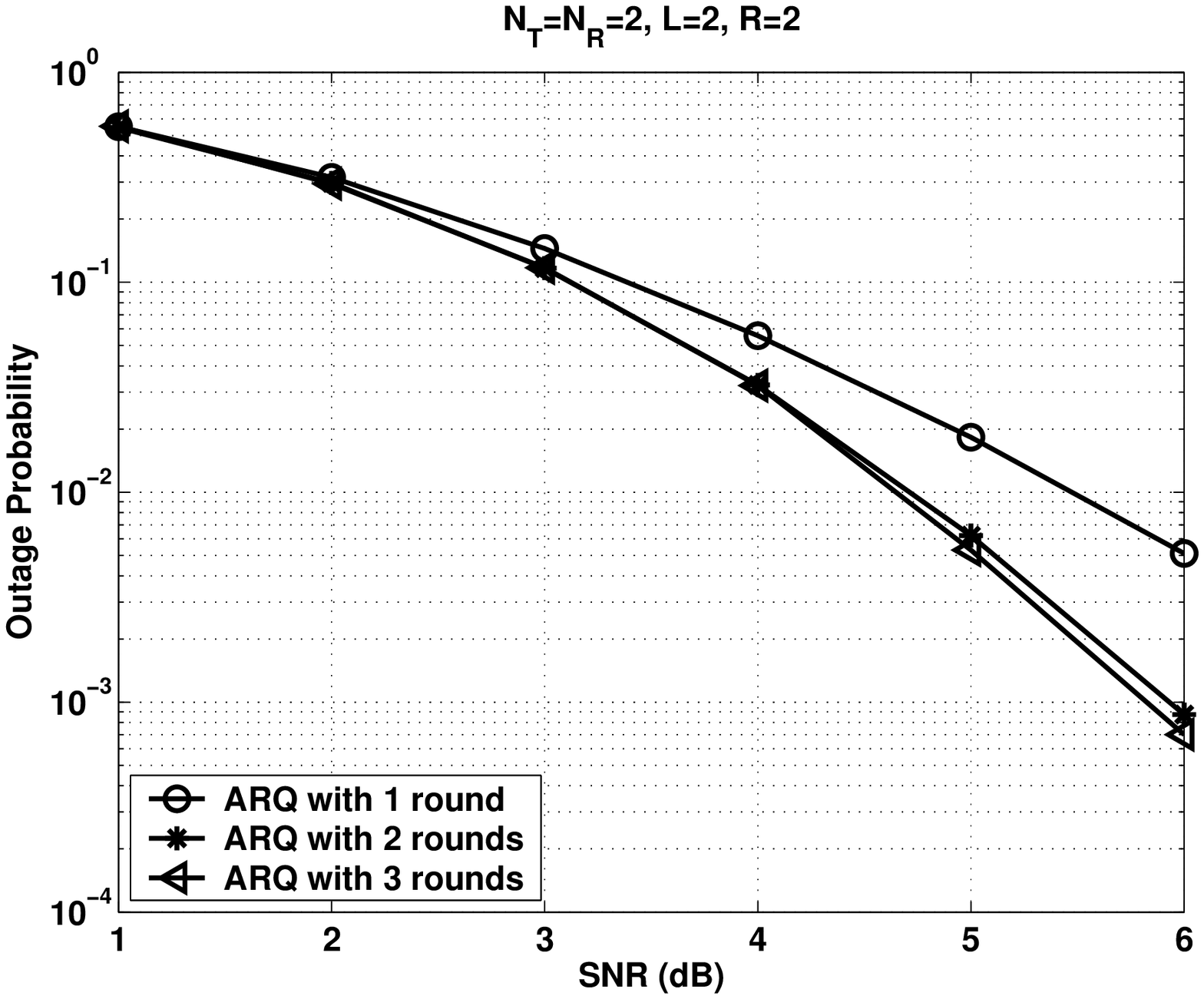}\hfill{}\includegraphics[width=7.5cm,height=9cm]{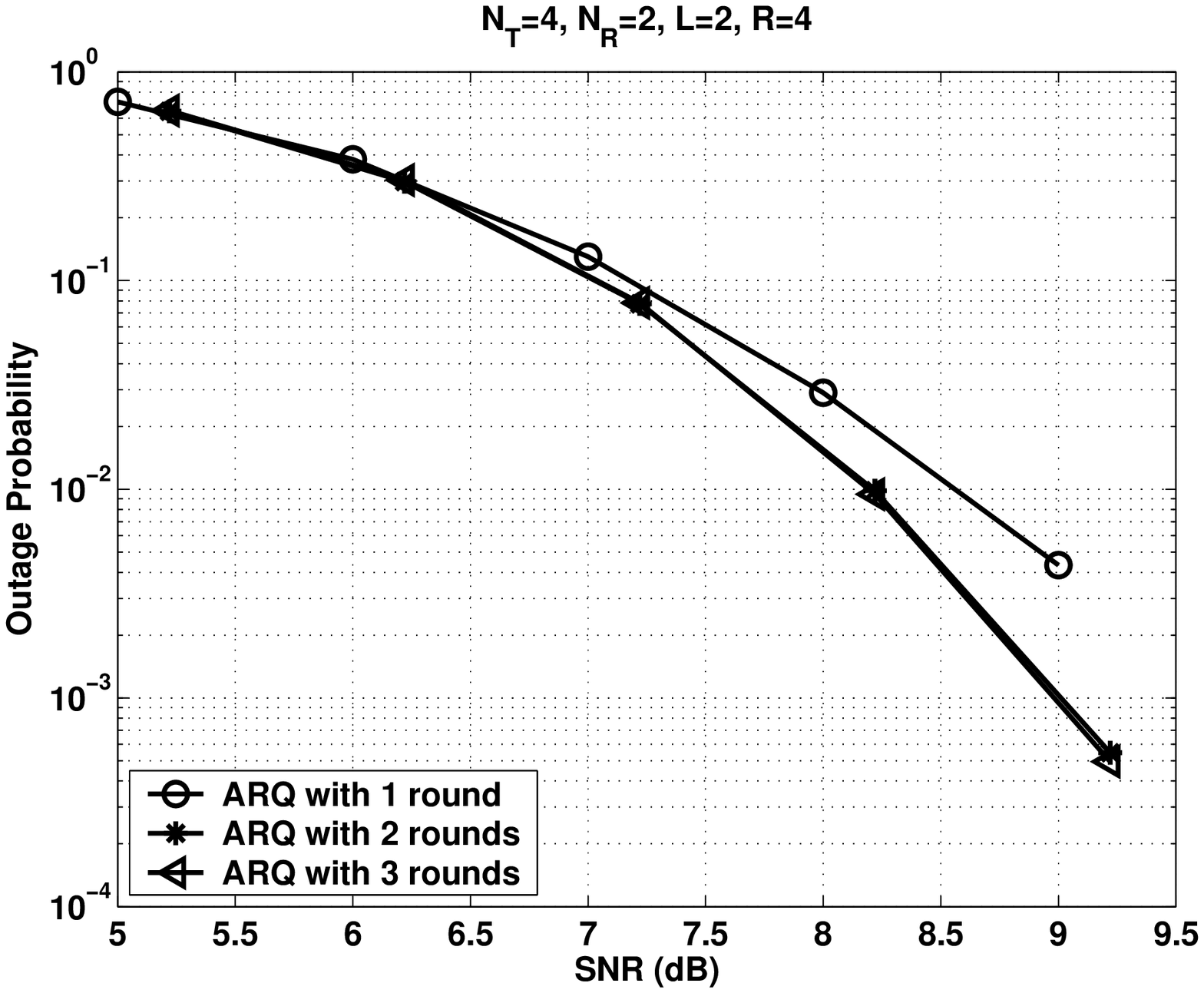}\hfill{}
\par\end{raggedright}

{\small ~~~~~~~~~~~~~~~~~~~~~~~~~~~~~~~~~~~~~~(a)}\hfill{}{\small (b)~~~~~~~~~~~~~~~~~~~~~~~~~~~~~~~~~~}{\small \par}

\raggedright{}\caption{\label{fig:Outage_Probability}Outage probabilty versus the maximum
number of rounds $K$ for $L=2$ taps, $N_{R}=2$, and: (a) $N_{T}=2$,
$R=2$, (b) $N_{T}=4$, $R=4$}

\end{figure*}

In Fig. \ref{fig:Outage_Probability}. a, we plot the outage probability
as a function of the ARQ delay $K$ for the two path MIMO-ISI channel
with two transmit and two receive antennas $\left(N_{T}=N_{R}=2\right)$,
and a target rate $R=2$. The ARQ diversity gain, due to the short-term
static channel dynamic, clearly appears when $K=2$. For instance,
a gain of approximately $1$dB is achieved at $5*10^{-3}$ outage
compared with the case of $K=1$ (i.e., no ARQ). When $K=3$, the
outage probability performance is similar to that of $K=2$. Fig.
\ref{fig:Outage_Probability}. b, shows the outage curves for $N_{T}=4$
and $N_{R}=2$ with a target rate $R=4$. We notice that as in the
previous configuration, $K=2$ and $K=3$ have the same outage performance,
while the overall diversity gain is more important than that corresponding
to $N_{T}=N_{R}=2$ (i.e., outage curve slopes are steeper than those
of the first configuration). Note that the stacking procedure (\ref{eq:Comb_RxSig_MAP})
relative to the optimal MAP-based turbo combiner creates $kN_{R}$
virtual receive antennas after $k$ rounds, but not all these virtual
antennas will translate into a receive diversity, because the target
rate $R$ has to be maintained as it can be seen from the expression
of the achievable information rate in (\ref{eq:Outage_eq2}). This
justifies the outage performance saturation after $K=2$. This issue
was recently addressed in \cite{DMD_Gamal_IT_2006} for MIMO ARQ with
flat fading, and it was demonstrated that the diversity gain does
not linearly increase with increase of the ARQ delay $K$. %
\footnote{In \cite[Theorem 2]{DMD_Gamal_IT_2006}, the authors demonstrated
that for the case of a short-term static flat fading MIMO ARQ channel,
the optimal diversity gain is $d^{*}\left(r_{e},K\right)=Kf\left(\frac{r_{e}}{K}\right)\,\,0\leq r_{e}<\min\left\{ N_{T},N_{R}\right\} $,
where $r_{e}$ is the multiplexing gain and $f$ is the piecewise
linear function connecting the points $\left(x,\left(N_{T}-x\right)\left(N_{R}-x\right)\right)$
for $x=0,\ldots,\min\left\{ N_{T},N_{R}\right\} $.%
}

\begin{figure}[t]
\noindent \begin{centering}
\includegraphics[scale=0.5]{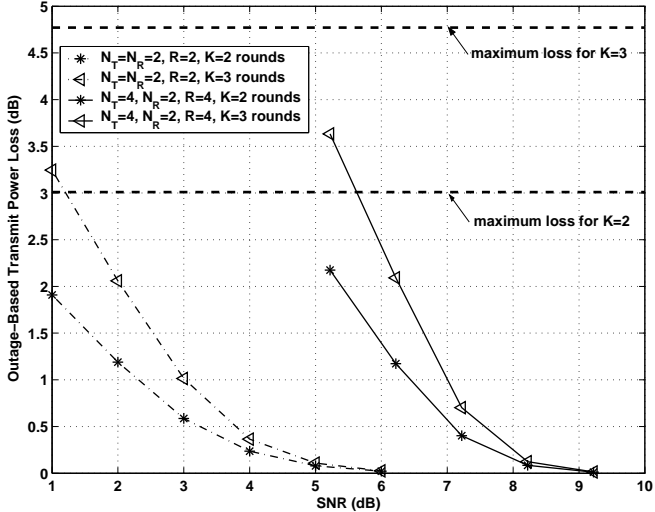}
\par\end{centering}

\caption{\label{fig:Outage_Tx_Pw_Loss}Outage-based transmit power loss for
$N_{T}=N_{R}=2$, $R=2$, and $N_{T}=4$, $N_{R}=2$, $R=4$}

\end{figure}

In Fig. \ref{fig:Outage_Tx_Pw_Loss}, we present the outage-based
transmit power loss for the considered MIMO configurations. We observe
that in the region of low SNR, the outage-based loss is significant
for both $K=2$ and $K=3$. When the outage probability is below $<10^{-2}$
(the region corresponding to FER values typically required in practical
systems), the transmit power loss is below $0.25$dB. This indicates
that in the corresponding SNR region, blocks are mainly error-free
during the first transmission, and only a small number of frames require
additional rounds. 

Motivated by these theoretical results, in the next section we design
a class of reduced complexity MMSE-based turbo combiners.

\section{Low Complexity MMSE-Based Turbo Packet Combining \label{sec:Proposed_Combiners}}

It is obvious that the complexity of the MAP turbo combining technique
presented in Subsection \ref{sub:GeneralArch_MAPcomb} is exponential
in the number of transmit antennas and channel use. In this section,
we introduce two low-complexity turbo packet combining techniques
using the MMSE criterion, and analyze their computational cost and
memory requirements.

\subsection{Signal-Level Turbo Combining\label{sub:Signal-Level-Turbo-Combining}}

Let us recall the MAP turbo combiner block communication model (\ref{eq:CommModel_OptimalMAP})
with a block length $\kappa=\kappa_{1}+\kappa_{2}+1\ll T$, where
$\kappa_{1}$ and $\kappa_{2}$ are the lengths of the forward and
backward filters, respectively. The corresponding $kN_{R}\kappa\times N_{T}\left(\kappa+L-1\right)$
sliding-window (around channel use $i$) communication model after
$k$ rounds is similar to (\ref{eq:CommModel_OptimalMAP}), and is
given as,%
\begin{equation}
\underline{\underline{\mathbf{y}}}_{i}^{\left(k\right)}=\underline{\underline{\mathbf{H}}}^{\left(k\right)}\underline{\mathbf{s}}_{i}+\underline{\underline{\mathbf{n}}}_{i}^{\left(k\right)},\label{eq:Slid_Model_SigLevel}\end{equation}
where \begin{equation}
\underline{\underline{\mathbf{y}}}_{i}^{\left(k\right)}\triangleq\left[\mathbf{y}_{i+\kappa_{1}}^{\left(1\right)^{\top}},\cdots,\mathbf{y}_{i+\kappa_{1}}^{\left(k\right)^{\top}},\cdots,\mathbf{y}_{i-\kappa_{2}}^{\left(1\right)^{\top}},\cdots,\mathbf{y}_{i-\kappa_{2}}^{\left(k\right)^{\top}}\right]^{\top}\label{eq:RxSig_SigLevel}\end{equation}
\begin{equation}
\underline{\underline{\mathbf{n}}}_{i}^{\left(k\right)}\triangleq\left[\mathbf{n}_{i+\kappa_{1}}^{\left(1\right)^{\top}},\cdots,\mathbf{n}_{i+\kappa_{1}}^{\left(k\right)^{\top}},\cdots,\mathbf{n}_{i-\kappa_{2}}^{\left(1\right)^{\top}},\cdots,\mathbf{n}_{i-\kappa_{2}}^{\left(k\right)^{\top}}\right]^{\mathrm{\top}}\label{eq:Noise_SigLevel}\end{equation}
are $kN_{R}\kappa\times1$ complex vectors, \begin{equation}
\underline{\mathbf{s}}_{i}\triangleq\left[\mathbf{s}_{i+\kappa_{1}}^{\top},\cdots,\mathbf{s}_{i-\kappa_{2}-L+1}^{\mathrm{\top}}\right]^{\top}\in\mathcal{S}^{N_{T}\left(\kappa+L-1\right)},\label{eq:SymblVec_SigLevel}\end{equation}
and $\underline{\underline{\mathbf{H}}}^{\left(k\right)}\in\mathbb{C}^{kN_{R}\kappa\times N_{T}\left(\kappa+L-1\right)}$
is defined similarly to (\ref{eq:BlockMat_MAP}). 

To compute, at the $n$th iteration extrinsic information $\phi_{m,t,i,n}^{e}$
about bit $b_{m,t,i}$, using signals received during rounds $1,\cdots,k$,
we jointly (over all rounds) cancel soft ISI in a parallel interference
cancellation (PIC) fashion. This yields a soft ISI-free signal vector
$\tilde{\underline{\underline{\mathbf{y}}}}_{i\mid\left(t,n\right)}^{\left(k\right)}\in\mathbb{C}^{kN_{R}\kappa}$
expressed as,\begin{equation}
\tilde{\underline{\underline{\mathbf{y}}}}_{i\mid\left(t,n\right)}^{\left(k\right)}\triangleq\underline{\underline{\mathbf{y}}}_{i}^{\left(k\right)}-\underline{\underline{\mathbf{H}}}^{\left(k\right)}\underline{\tilde{\mathbf{s}}}_{i\mid\left(t,n\right)},\label{eq:CondInterfrenceCancel_SigLevel}\end{equation}
where $\underline{\tilde{\mathbf{s}}}_{i\mid\left(t,n\right)}$ is
the conditional average of symbol vector $\underline{\mathbf{s}}_{i}$
with zero at the $\left(\kappa_{1}N_{T}+t\right)$th position,\begin{equation}
\underline{\tilde{\mathbf{s}}}_{i\mid\left(t,n\right)}\triangleq\mathbb{{E}}\left[\underline{\mathbf{s}}_{i}\mid\phi_{m',t',i',n}^{a}:\left(t',i'\right)\neq\left(t,i\right)\right].\label{eq:Conf_SymblVector_SigLevel}\end{equation}
The components of $\tilde{\underline{\underline{\mathbf{y}}}}_{i\mid\left(t,n\right)}^{\left(k\right)}$
are then combined using an unconditional MMSE filter to produce the
scalar input $\xi_{t,i,n}^{\left(k\right)}$ for the soft demapper.
Applying the matrix inversion lemma \cite{Haykin} similarly to \cite[eq. 6]{Tuchler_TSP_2002},
we can write the output of the unconditional MMSE filter as, 

\begin{equation}
\xi_{t,i,n}^{\left(k\right)}=\zeta_{t,n}^{\left(k\right)}\mathbf{e}_{t}^{\top}\underline{\underline{\mathbf{H}}}^{\left(k\right)^{H}}\mathbf{A}_{n}^{\left(k\right)^{-1}}\tilde{\underline{\underline{\mathbf{y}}}}_{i\mid\left(t,n\right)}^{\left(k\right)},\label{eq:UncondMMSE_Filtering_SigLevel}\end{equation}
 where\begin{equation}
\mathbf{A}_{n}^{\left(k\right)}=\underline{\underline{\mathbf{H}}}^{\left(k\right)}\mathbf{\Xi}_{n}\underline{\underline{\mathbf{H}}}^{\left(k\right)^{H}}+\sigma^{2}\mathbf{I}_{kN_{R}\kappa}\,\in\mathbb{C}^{kN_{R}\kappa\times kN_{R}\kappa},\label{eq:MatrixAk_uncondMMSE_SigLevel}\end{equation}

\begin{equation}
\mathbf{\Xi}_{n}=\mathbf{I}_{\kappa+L-1}\otimes\mathbf{\tilde{\Xi}}_{n}\in\mathbb{C}^{N_{T}\left(\kappa+L-1\right)\times N_{T}\left(\kappa+L-1\right)},\label{eq:SymbolCovMat_SigLevel}\end{equation}
\begin{equation}
\mathbf{\tilde{\Xi}}_{n}\triangleq\mathrm{diag}\left\{ \tilde{\sigma}_{1,n}^{2},\cdots,\tilde{\sigma}_{N_{T},n}^{2}\right\} ,\label{eq:SymbolCovMat_SigLevel2}\end{equation}
\begin{equation}
\mathbf{e}_{t}\triangleq\left[\underbrace{0,\cdots,0}_{\kappa_{1}N_{T}+t-1},1,\underbrace{0,\cdots,0}_{\left(\kappa_{2}+L\right)N_{T}-t}\right]^{\top}\in\mathbb{C}^{N_{T}\left(\kappa+L-1\right)},\label{eq:canocialBasisvector_SigLevel}\end{equation}
\begin{equation}
\zeta_{t,n}^{\left(k\right)}=\left(1+\left(1-\tilde{\sigma}_{t,n}^{2}\right)\mathbf{e}_{t}^{\top}\underline{\underline{\mathbf{H}}}^{\left(k\right)^{H}}\mathbf{A}_{n}^{\left(k\right)^{-1}}\underline{\underline{\mathbf{H}}}^{\left(k\right)}\mathbf{e}_{t}\right)^{-1},\label{eq:Scalar_term_MMSEfilter_SigLevel}\end{equation}
 and $\tilde{\sigma}_{t,n}^{2}$ is the unconditional variance at
iteration $n$ of symbols $\left\{ s_{t,i}\right\} _{i=0}^{T-1}$
transmitted over antenna $t$,\begin{equation}
\tilde{\sigma}_{t,n}^{2}=\frac{1}{T}\sum_{i=0}^{T-1}\mathbb{{E}}\left[\left|s_{t,i}-\tilde{s}_{t,i,n}\right|^{2}\mid\phi_{m,t,i,n}^{a}:m=1,\cdots,M\right],\label{eq:UncondSymbolVariance}\end{equation}
\begin{equation}
\tilde{s}_{t,i,n}\triangleq\mathbb{{E}}\left[s_{t,i}\mid\phi_{m,t,i,n}^{a}:m=1,\cdots,M\right]\label{eq:CondSymbolAverage}\end{equation}
 is the conditional average of symbol $s_{t,i}$ at iteration $n$.

Combining the soft PIC (\ref{eq:CondInterfrenceCancel_SigLevel})
and unconditional MMSE filtering (\ref{eq:UncondMMSE_Filtering_SigLevel})
steps, and after some matrix manipulations, we can write the soft
demapper input $\xi_{t,i,n}^{\left(k\right)}$ as,

\begin{equation}
\xi_{t,i,n}^{\left(k\right)}=\mathbf{F}_{t,n}^{\left(k\right)}\underline{\mathbf{z}}_{i}^{\left(k\right)}-\mathbf{B}_{t,n}^{\left(k\right)}\underline{\tilde{\mathbf{s}}}_{i\mid\left(t,n\right)}.\label{eq:Forward_Backward_SigLevel}\end{equation}
$\mathbf{F}_{t,n}^{\left(k\right)}$ and $\mathbf{B}_{t,n}^{\left(k\right)}$
are the forward and backward filters corresponding to antenna $t$
at the $n$th iteration,\begin{figure*}[!t]
\normalsize 
\setcounter{mytempeqncnt}{\value{equation}} 
\setcounter{equation}{44}
\begin{equation} \phi_{m,t,i,n}^{e^{\left[Sig\right]}}=\log\frac{{\displaystyle \sum_{s\in\mathcal{S}_{m}^{1}}}\exp\left\{ -\frac{1}{2\delta_{t,n}^{\left(k\right)^{2}}}\left|\xi_{t,i,n}^{\left(k\right)}-\alpha_{t,n}^{\left(k\right)}s\right|^{2}+\sum_{m'\neq m}\varphi_{m'}^{-1}(s)\phi_{m',t,i,n}^{a}\right\} }{{\displaystyle \sum_{\mathbf{s}\in\mathcal{S}_{m}^{0}}}\exp\left\{ -\frac{1}{2\delta_{t,n}^{\left(k\right)^{2}}}\left|\xi_{t,i,n}^{\left(k\right)}-\alpha_{t,n}^{\left(k\right)}s\right|^{2}+\sum_{m'\neq m}\varphi_{m'}^{-1}(s)\phi_{m',t,i,n}^{a}\right\} },\label{eq:Ext_Info_SigLevel}\end{equation}
\setcounter{equation}{35} 
\hrulefill 
\vspace*{4pt}
\end{figure*}\begin{figure*}[!t]
\normalsize 
\setcounter{mytempeqncnt}{\value{equation}} 
\setcounter{equation}{46}
\begin{equation} \phi_{m,t,i,n}^{e^{\left[Symb\right]}}=\log\frac{{\displaystyle \sum_{s\in\mathcal{S}_{m}^{1}}}\exp\left\{ -\frac{1}{2}\left(\boldsymbol{{\breve{\xi}}}_{t,i,n}^{\left(k\right)}-s\boldsymbol{{\breve{\alpha}}}_{t,n}^{\left(k\right)}\right)^{H}\boldsymbol{{\Delta}}_{t,n}^{\left(k\right)^{-1}}\left(\boldsymbol{{\breve{\xi}}}_{t,i,n}^{\left(k\right)}-s\boldsymbol{{\breve{\alpha}}}_{t,n}^{\left(k\right)}\right)+\sum_{m'\neq m}\varphi_{m'}^{-1}(s)\phi_{m',t,i,n}^{a}\right\} }{{\displaystyle \sum_{\mathbf{s}\in\mathcal{S}_{m}^{0}}}\exp\left\{ -\frac{1}{2}\left(\boldsymbol{{\breve{\xi}}}_{t,i,n}^{\left(k\right)}-s\boldsymbol{{\breve{\alpha}}}_{t,n}^{\left(k\right)}\right)^{H}\boldsymbol{{\Delta}}_{t,n}^{\left(k\right)^{-1}}\left(\boldsymbol{{\breve{\xi}}}_{t,i,n}^{\left(k\right)}-s\boldsymbol{{\breve{\alpha}}}_{t,n}^{\left(k\right)}\right)+\sum_{m'\neq m}\varphi_{m'}^{-1}(s)\phi_{m',t,i,n}^{a}\right\} },\label{eq:Ext_Info_FilterLevel}\end{equation}
\setcounter{equation}{35} 
\hrulefill 
\vspace*{4pt}
\end{figure*}

\begin{equation}
\mathbf{F}_{t,n}^{\left(k\right)}=\left(\sigma^{2}+\left(1-\tilde{\sigma}_{t,n}^{2}\right)\mathbf{e}_{t}^{\top}\mathbf{\mathbf{\Lambda}}_{n}^{\left(k\right)}\mathbf{\Upsilon}^{\left(k\right)}\mathbf{e}_{t}\right)^{-1}\mathbf{e}_{t}^{\top}\mathbf{\mathbf{\Lambda}}_{n}^{\left(k\right)},\label{eq:Forward_SigLevel}\end{equation}

\noindent \begin{flushleft}
\begin{equation}
\mathbf{B}_{t,n}^{\left(k\right)}=\mathbf{F}_{t,n}^{\left(k\right)}\mathbf{\Upsilon}^{\left(k\right)}.\label{eq:Backward_SigLevel}\end{equation}

\par\end{flushleft}

\noindent \begin{flushleft}
$\mathbf{\mathbf{\Lambda}}_{n}^{\left(k\right)}$, $\underline{\mathbf{z}}_{i}^{\left(k\right)}$,
and $\mathbf{\Upsilon}^{\left(k\right)}$are given as 
\par\end{flushleft}

\noindent \begin{flushleft}
\begin{equation}
\mathbf{\mathbf{\Lambda}}_{n}^{\left(k\right)}=\mathbf{I}_{N_{T}\left(\kappa+L-1\right)}-\mathbf{\Upsilon}^{\left(k\right)}\left(\mathbf{\Upsilon}^{\left(k\right)}+\sigma^{2}\mathbf{\Xi}_{n}^{-1}\right)^{-1},\label{eq:Lambda_n_SigLevel}\end{equation}

\par\end{flushleft}

\noindent \begin{flushleft}
\begin{equation}
\begin{cases}
\underline{\mathbf{z}}_{i}^{\left(k\right)} & =\underline{\mathbf{z}}_{i}^{\left(k-1\right)}+\underline{\mathbf{H}}^{\left(k\right)^{H}}\underline{\mathbf{y}}_{i}^{\left(k\right)}\\
\underline{\mathbf{z}}_{i}^{\left(0\right)} & =\mathbf{0}_{N_{T}\left(\kappa+L-1\right)\times1},\end{cases}\label{eq:Recursion1_SigLevel}\end{equation}

\par\end{flushleft}

\noindent \begin{flushleft}
\begin{equation}
\begin{cases}
\mathbf{\Upsilon}^{\left(k\right)} & =\mathbf{\Upsilon}^{\left(k-1\right)}+\underline{\mathbf{H}}^{\left(k\right)^{H}}\underline{\mathbf{H}}^{\left(k\right)}\\
\mathbf{\Upsilon}^{\left(0\right)} & =\mathbf{0}_{N_{T}\left(\kappa+L-1\right)\times N_{T}\left(\kappa+L-1\right)}.\end{cases}\label{eq:Recursion2_SigLevel}\end{equation}

\par\end{flushleft}

\noindent $\underline{\mathbf{H}}^{\left(k\right)}\in\mathbb{C}^{N_{R}\kappa\times N_{T}\left(\kappa+L-1\right)}$
and $\underline{\mathbf{y}}_{i}^{\left(k\right)}$ are the block Toeplitz
matrix and signal output of the sliding-window communication model
at round $k$, respectively, and are given as, 

\noindent \begin{equation}
\underline{\mathbf{H}}^{\left(k\right)}\triangleq\left[\begin{array}{ccccc}
\mathbf{H}_{0}^{\left(k\right)} & \cdots & \mathbf{H}_{L-1}^{\left(k\right)}\\
 & \ddots &  & \ddots\\
 &  & \mathbf{H}_{0}^{\left(k\right)} & \cdots & \mathbf{H}_{L-1}^{\left(k\right)}\end{array}\right],\label{eq:BlockToeplitz_round_k}\end{equation}

\noindent \begin{equation}
\underline{\mathbf{y}}_{i}^{\left(k\right)}\triangleq\left[\mathbf{y}_{i+\kappa_{1}}^{\left(k\right)^{\top}},\cdots,\mathbf{y}_{i-\kappa_{2}}^{\left(k\right)^{\top}}\right]^{\top}\in\mathbb{C}^{N_{R}\kappa},\label{eq:SlidWin_Out_round_k}\end{equation}

\noindent \begin{equation}
\underline{\mathbf{y}}_{i}^{\left(k\right)}=\underline{\mathbf{H}}^{\left(k\right)}\underline{\mathbf{s}}_{i}+\underline{\mathbf{n}}_{i}^{\left(k\right)},\label{eq:SlidWin_CommModel_round_k}\end{equation}

\noindent \begin{equation}
\underline{\mathbf{n}}_{i}^{\left(k\right)}\triangleq\left[\mathbf{n}_{i+\kappa_{1}}^{\left(k\right)^{\top}},\cdots,\mathbf{n}_{i-\kappa_{2}}^{\left(k\right)^{\top}}\right]^{\top}\in\mathbb{C}^{N_{R}\kappa}.\label{eq:Slid_Win_Noise_round_k}\end{equation}

\noindent Recursions (\ref{eq:Recursion1_SigLevel}) and (\ref{eq:Recursion2_SigLevel})
are easily obtained by invoking (\ref{eq:RxSig_SigLevel}) and the
general structure (\ref{eq:BlockMat_MAP}). Details about the derivation
of (\ref{eq:Forward_Backward_SigLevel}) are omitted because of space
limitation. Assuming the conditional soft demapper input is Gaussian,
i.e., $\left(\xi_{t,i,n}^{\left(k\right)}\mid s_{t,i}\right)\sim\mathcal{N}\left(\alpha_{t,n}^{\left(k\right)},\delta_{t,n}^{\left(k\right)^{2}}\right)$,
extrinsic information $\phi_{m,t,i,n}^{e^{\left[Sig\right]}}$ can
be computed according to (\ref{eq:Ext_Info_SigLevel}), where \setcounter{equation}{45}

\begin{equation}
\begin{cases}
\alpha_{t,n}^{\left(k\right)} & =\mathbf{B}_{t,n}^{\left(k\right)}\mathbf{e}_{t}\\
\delta_{t,n}^{\left(k\right)^{2}} & =\left(1-\alpha_{t,n}^{\left(k\right)}\right)\alpha_{t,n}^{\left(k\right)},\end{cases}\label{eq:Ext_Info_Param_Gauss_SigLevel}\end{equation}
and $\mathcal{S}_{m}^{b}=\left\{ s\in\mathcal{S}\mid\varphi_{m}^{-1}(s)=b\right\} $.
The signal-level combining algorithm is summarized in Table \ref{tab:SummarySigLevel}.%
\begin{table*}
\centering{}\caption{\label{tab:SummarySigLevel}Summary of the signal-level turbo packet
combining algorithm}
\begin{tabular}{lllll}
\hline 
 & \multicolumn{4}{l}{}\tabularnewline
\textbf{\small 0.} & \multicolumn{4}{l}{\textbf{\small Initialization}}\tabularnewline
 & \multicolumn{4}{l}{{\small Initialize $\mathbf{\Upsilon}^{\left(0\right)}$ and $\left\{ \underline{\mathbf{z}}_{i}^{\left(0\right)}\right\} _{i=0}^{T-1}$
with $\mathbf{0}_{N_{T}\left(\kappa+L-1\right)}$ and vectors $\mathbf{0}_{N_{T}\left(\kappa+L-1\right)\times1}$,
respectively.}}\tabularnewline
\textbf{\small 1.} & \multicolumn{4}{l}{\textbf{\small Combining at round $k$}}\tabularnewline
 & \textbf{\small 1.1.} & \multicolumn{3}{l}{{\small Update $\left\{ \underline{\mathbf{z}}_{i}^{\left(k\right)}\right\} _{i=0}^{T-1}$
and $\mathbf{\Upsilon}^{\left(k\right)}$ according to (\ref{eq:Recursion1_SigLevel})
and (\ref{eq:Recursion2_SigLevel}).}}\tabularnewline
 & \textbf{\small 1.2.} & \multicolumn{3}{l}{{\small For~ $n=1,\cdots,N$}}\tabularnewline
 &  & \textbf{\small 1.2.1.} & \multicolumn{2}{l}{{\small Compute: conditional symbol averages and unconditional variances
using (\ref{eq:CondSymbolAverage}) and (\ref{eq:UncondSymbolVariance}).}}\tabularnewline
 &  & \textbf{\small 1.2.2.} & \multicolumn{2}{l}{{\small Compute: $\mathbf{\mathbf{\Lambda}}_{n}^{\left(k\right)}$
using (\ref{eq:Lambda_n_SigLevel}).}}\tabularnewline
 &  & \textbf{\small 1.2.3.} & \multicolumn{2}{l}{{\small For ~$t=1,\cdots,N_{T}$}}\tabularnewline
 &  &  & \textbf{\small 1.2.3.1.} & {\small Compute: $\mathbf{F}_{t,n}^{\left(k\right)}$, $\mathbf{B}_{t,n}^{\left(k\right)}$,
$\alpha_{t,n}^{\left(k\right)}$, and $\delta_{t,n}^{\left(k\right)^{2}}$
using (\ref{eq:Forward_SigLevel}), (\ref{eq:Backward_SigLevel}),
and (\ref{eq:Ext_Info_Param_Gauss_SigLevel}).}\tabularnewline
 &  &  & \textbf{\small 1.2.3.2.} & {\small For each~ $i=0,\cdots,T-1$, compute the soft demapper input
$\xi_{t,i,n}^{\left(k\right)}$ according to (\ref{eq:Forward_Backward_SigLevel}). }\tabularnewline
 &  &  & \textbf{\small 1.2.3.3.} & {\small For each $m=1,\cdots,M$, compute extrinsic information $\phi_{m,t,i,n}^{e^{\left[Sig\right]}}$
using (\ref{eq:Ext_Info_SigLevel}).}\tabularnewline
 &  & \textbf{\small 1.2.4.} & \multicolumn{2}{l}{{\small End }\textbf{\small 1.2.3.}}\tabularnewline
 & \textbf{\small 1.3.} & \multicolumn{3}{l}{{\small End }\textbf{\small 1.2.}}\tabularnewline
 &  &  &  & \tabularnewline
\hline
\end{tabular}
\end{table*}

Note that the forward-backward filtering structure (\ref{eq:Forward_Backward_SigLevel})
together with recursions (\ref{eq:Recursion1_SigLevel}) and (\ref{eq:Recursion2_SigLevel})
present the core part of the proposed algorithm, and allow a reduced
computational complexity and an optimized implementation. Indeed,
equations (\ref{eq:Recursion1_SigLevel}) and (\ref{eq:Recursion2_SigLevel})
allow to use at each ARQ round all signals and channel matrices corresponding
to previous rounds $k-1,\cdots,1$ without being required to be explicitly
stored in the receiver. This is performed in a recursive fashion using
modified versions of the sliding window input and matrix ( i.e., $\underline{\mathbf{H}}^{\left(k\right)^{H}}\underline{\mathbf{y}}_{i}^{\left(k\right)}$
and $\underline{\mathbf{H}}^{\left(k\right)^{H}}\underline{\mathbf{H}}^{\left(k\right)}$,
respectively) at round $k$.

\subsection{Symbol-Level Turbo Combining}

In this combining scheme, we propose to perform equalization separately
for each round $k$ based on the communication model (\ref{eq:SlidWin_CommModel_round_k}).
Then, soft combining is conducted at the level of unconditional MMSE
filter outputs: The output at iteration $n$ of round $k$ is combined
with the outputs obtained at the last iteration of previous rounds
$k-1,\cdots,1$. As in the previous subsection, let $\breve{\xi}_{t,i,n}^{\left(k\right)}$
denote the filter output %
\footnote{The forward and backward filters can be easily derived using the equations
in the previous subsection and assuming $k=1$. %
} at iteration $n$ of round $k$, and $\left(\breve{\xi}_{t,i,n}^{\left(k\right)}\mid s_{t,i}\right)\sim\mathcal{N}\left(\breve{\alpha}_{t,n}^{\left(k\right)},\breve{\delta}_{t,n}^{\left(k\right)^{2}}\right)$.
The soft demapper, which has a vector input in this case, computes
extrinsic information $\phi_{m,t,i,n}^{e^{\left[Symb\right]}}$ according
to (\ref{eq:Ext_Info_FilterLevel}), where \setcounter{equation}{47}

\noindent \begin{flushleft}
\begin{equation}
\boldsymbol{{\breve{\xi}}}_{t,i,n}^{\left(k\right)}\triangleq\left[\breve{\xi}_{t,i,N}^{\left(1\right)},\cdots,\breve{\xi}_{t,i,N}^{\left(k-1\right)},\breve{\xi}_{t,i,n}^{\left(k\right)}\right]^{\top}\in\mathbb{C}^{k},\label{eq:Filter_Outputs}\end{equation}

\par\end{flushleft}

\noindent \begin{flushleft}
\begin{equation}
\boldsymbol{{\breve{\alpha}}}_{t,n}^{\left(k\right)}\triangleq\left[\breve{\alpha}_{t,N}^{\left(1\right)},\cdots,\breve{\alpha}_{t,N}^{\left(k-1\right)},\breve{\alpha}_{t,n}^{\left(k\right)}\right]^{\top}\in\mathbb{C}^{k},\label{eq:Filter_Output_Gains}\end{equation}

\par\end{flushleft}

\noindent \begin{flushleft}
and $\boldsymbol{{\Delta}}_{t,n}^{\left(k\right)}$ is the covariance
matrix of $\left(\boldsymbol{{\breve{\xi}}}_{t,i,n}^{\left(k\right)}\mid s_{t,i}\right)$
which can be approximated as (assuming residual ISI plus noise terms
at different rounds are independent),\begin{equation}
\boldsymbol{{\Delta}}_{t,n}^{\left(k\right)}\approx\mathrm{diag}\left\{ \breve{\delta}_{t,N}^{\left(1\right)^{2}},\cdots,\breve{\delta}_{t,N}^{\left(k-1\right)^{2}},\breve{\delta}_{t,n}^{\left(k\right)^{2}}\right\} .\label{eq:Res_ISI_noise_CovMat}\end{equation}

\par\end{flushleft}

\noindent \begin{flushleft}
The algorithm is summarized in Table \ref{tab:SummarySymbLevel}.%
\begin{table*}
\centering{}\caption{\label{tab:SummarySymbLevel}Summary of the symbol-level turbo packet
combining algorithm}
\begin{tabular}{lllll}
\hline 
 & \multicolumn{4}{l}{}\tabularnewline
\textbf{\small 0.} & \multicolumn{4}{l}{\textbf{\small Initialization}}\tabularnewline
 & \multicolumn{4}{l}{{\small Initialize $\left\{ \boldsymbol{{\breve{\xi}}}_{t,i}\right\} _{i=0}^{T-1}$,
$\boldsymbol{{\breve{\alpha}}}_{t}$, and $\boldsymbol{{\breve{\delta}}}_{t}^{2}$
with empty vectors for $t=1,\cdots,N_{T}$.}}\tabularnewline
\textbf{\small 1.} & \multicolumn{4}{l}{\textbf{\small Combining at round $k$}}\tabularnewline
 & \textbf{\small 1.1.} & \multicolumn{3}{l}{{\small For~ $n=1,\cdots,N$}}\tabularnewline
 &  & \textbf{\small 1.1.1.} & \multicolumn{2}{l}{{\small Compute: conditional symbol averages and unconditional variances
using (\ref{eq:CondSymbolAverage}) and (\ref{eq:UncondSymbolVariance}).}}\tabularnewline
 &  & \textbf{\small 1.1.2.} & \multicolumn{2}{l}{{\small For ~$t=1,\cdots,N_{T}$}}\tabularnewline
 &  &  & \textbf{\small 1.1.2.1.} & {\small Compute: forward and backward filters, $\breve{\alpha}_{t,n}^{\left(k\right)}$,
and $\breve{\delta}_{t,n}^{\left(k\right)^{2}}$ as in Subsection
\ref{sub:Signal-Level-Turbo-Combining}.}\tabularnewline
 &  &  & \textbf{\small 1.1.2.2.} & {\small For each~ $i=0,\cdots,T-1$, compute the filter output $\breve{\xi}_{t,i,n}^{\left(k\right)}$.}\tabularnewline
 &  &  & \textbf{\small 1.1.2.3.} & {\small For each $m=1,\cdots,M$, compute extrinsic information $\phi_{m,t,i,n}^{e^{\left[Symb\right]}}$
using (\ref{eq:Ext_Info_FilterLevel}).}\tabularnewline
 &  & \textbf{\small 1.2.3.} & \multicolumn{2}{l}{{\small End }\textbf{\small 1.1.2.}}\tabularnewline
 & \textbf{\small 1.2.} & \multicolumn{3}{l}{{\small End }\textbf{\small 1.1.}}\tabularnewline
 & \textbf{\small 1.3.} & \multicolumn{3}{l}{{\small Update: $\left\{ \boldsymbol{{\breve{\xi}}}_{t,i}:=\left[\boldsymbol{{\breve{\xi}}}_{t,i}\,\,\breve{\xi}_{t,i,N}^{\left(k\right)}\right]\right\} _{i=0}^{T-1}$,
$\boldsymbol{{\breve{\alpha}}}_{t}:=\left[\boldsymbol{{\breve{\alpha}}}_{t}\,\,\breve{\alpha}_{t,N}^{\left(k\right)}\right]$,
and $\boldsymbol{{\breve{\delta}}}_{t}^{2}:=\left[\boldsymbol{{\breve{\delta}}}_{t}^{2}\,\,\breve{\delta}_{t,N}^{\left(k\right)^{2}}\right]$
for $t=1,\cdots,N_{T}$.}}\tabularnewline
 &  &  &  & \tabularnewline
\hline
\end{tabular}
\end{table*}
 
\par\end{flushleft}

\subsection{Complexity Analysis\label{sub:Complexity-Analysis}}

In this subsection, we focus on the analysis of the computational
cost of forward and backward filters as well as the memory requirements
for the proposed algorithms. The other steps are similar and have
the same complexity for both algorithms. We also provide comparisons
with the conventional LLR-level combining technique. 

In the case of signal-level turbo combining, the computation of forward
and backward filters involves, at each round $k$ and iteration $n$,
one inversion of a $N_{T}\left(\kappa+L-1\right)\times N_{T}\left(\kappa+L-1\right)$
matrix (i.e., matrix $\mathbf{\Upsilon}^{\left(k\right)}+\sigma^{2}\mathbf{\Xi}_{n}^{-1}$
in eq. (\ref{eq:Lambda_n_SigLevel})) for computing $\mathbf{\mathbf{\Lambda}}_{n}^{\left(k\right)}$,
and whose cost is $\mathcal{O}\left(N_{T}^{3}\kappa^{3}\right)$ (assuming
$\kappa\gg L$, and neglecting the cost of obtaining $\mathbf{\Xi}_{n}^{-1}=\mathbf{I}_{\kappa+L-1}\otimes\mathbf{\tilde{\Xi}}_{n}^{-1}$
since $\mathbf{\tilde{\Xi}}_{n}$ is diagonal). This indicates that
the computational complexity of the signal-level combining scheme
is less sensitive to $k$. The number of rounds only influences the
number of additions required for obtaining vectors $\left\{ \underline{\mathbf{z}}_{i}^{\left(k\right)}\right\} _{0\leq i\leq T-1}$
and matrix $\mathbf{\Upsilon}^{\left(k\right)}$ , according to (\ref{eq:Recursion1_SigLevel})
and (\ref{eq:Recursion2_SigLevel}), respectively. The cost of these
steps is 

\begin{equation}
\triangle N_{Add}=N_{T}^{2}\left(\kappa+L-1\right)^{2}+N_{R}\kappa T\label{eq:DeltaN_add}\end{equation}

\noindent for each round $k>1$. Note that the number of operations
required for obtaining $\underline{\mathbf{H}}^{\left(k\right)^{H}}\underline{\mathbf{H}}^{\left(k\right)}$
and $\underline{\mathbf{H}}^{\left(k\right)^{H}}\underline{\mathbf{y}}_{i}^{\left(k\right)}$
in not considered in (\ref{eq:DeltaN_add}) since symbol-level combining
also involves the same operations. Therefore, the computational cost
of forward and backward filters is almost the same for both combining
algorithms. Note that the significant reduction in the complexity
of the signal-level combining scheme (with respect to the dimensionality
of the sliding-window model (\ref{eq:Slid_Model_SigLevel}) used by
the algorithm) is due to recursion (\ref{eq:Recursion2_SigLevel})
which consists of writing $\underline{\underline{\mathbf{H}}}^{\left(k\right)^{H}}\underline{\underline{\mathbf{H}}}^{\left(k\right)}$
as the sum $\sum_{u=1}^{k}\underline{\mathbf{H}}^{\left(u\right)^{H}}\underline{\mathbf{H}}^{\left(u\right)}$.

Memory requirements for the two proposed schemes are determined by
the update steps Tables \ref{tab:SummarySigLevel}. \textbf{1.1} and
\ref{tab:SummarySymbLevel}. \textbf{1.3}. For the signal-level combining
technique, a $N_{T}\left(\kappa+L-1\right)\times N_{T}\left(\kappa+L-1\right)$
complex matrix is required to accumulate channel matrices $\underline{\mathbf{H}}^{\left(k\right)^{H}}\underline{\mathbf{H}}^{\left(k\right)}$
according to (\ref{eq:Recursion2_SigLevel}) (and therefore generating
$\mathbf{\Upsilon}^{\left(k\right)}$), in addition to a $N_{T}\left(\kappa+L-1\right)\times T$
complex matrix that serves to accumulate signal vectors $\left\{ \underline{\mathbf{z}}_{i}^{\left(k\right)}\right\} _{i=0}^{T-1}$
using (\ref{eq:Recursion1_SigLevel}). Note that these two recursions,
i.e., (\ref{eq:Recursion1_SigLevel}) and (\ref{eq:Recursion2_SigLevel}),
avoid the storage of all signals and channel matrices as in MAP turbo
combining. In the case of symbol-level combining, only $N_{T}$ complex
matrices of size $K\times T$ and two $K\times N_{T}$ complex matrices
are required to store filter outputs and their corresponding parameters,
i.e., symbol gains and residual ISI plus thermal noise variances.
Therefore, signal-level combining requires slightly more memory than
its symbol-level counterpart, because only two or three ARQ rounds
are considered (according to the outage analysis in Subsection \ref{sub:Outage-Analysis})
and in general $\kappa\gg L$.

Finally, note that in the case of conventional LLR-level combining,
soft equalization is separately performed for each ARQ round exactly
as in symbol-level combining, while extrinsic LLRs are added together
before decoding. This translates into $N_{T}MTN$ real additions at
each round, and a real vector of size $N_{T}MT$ to combine extrinsic
values. Therefore, the three combining strategies have similar implementation
requirements. They slightly differ in the number of additions and
storage memory.

\section{Numerical Results\label{sec:Numerical-Results}}

In this section, we provide simulated BLER and throughput performance
for the proposed turbo packet combining techniques presented in Section
\ref{sec:Proposed_Combiners}. Considering some representative MIMO
configurations, our main focus is to demonstrate that the signal-level
turbo combining approach has better ISI cancellation capability and
diversity gain than the symbol-level approach. We also show that both
techniques provide better performance than conventional LLR-level
combining.

\subsection{Simulation Settings}

In all simulations, we use an ST-BICM scheme composed of a $64$-state
$\frac{1}{2}$-rate convolutional code with polynomial generators
$\left(133_{8},171_{8}\right)$. The length of the code frame is $1800$
bits including tail bits. We consider either quadrature phase shift
keying (QPSK) or 16-state quadrature amplitude modulation (QAM) depending
on the target rate $R$ of the ST-BICM code. The MIMO-ISI channel
has the same profile as in Subsection \ref{sub:Outage-Analysis},
i.e., two equal power taps. With respect to the outage analysis in
Section \ref{sec:OptimalPacketComb_and_Outage}, we consider a ARQ
delay $K=2$. We verified, with simulations, that for the considered
ST-BICM code, the improvement in BLER performance is only incremental
when $K>2$. Note that in \cite{Ait-Idir_WCNC_08}, only a four-state
code is used, and performance results are reported with a maximum
number of rounds $K=3$. Simulations are carried out as in Subsection
\ref{sub:Outage-Analysis}, i.e., the transmission of an information
block is stopped and the system moves on to the next block when an
ACK message is received or the decoding outcome is erroneous after
round $K=2$. 

Note that the benefits of an ARQ mechanism appear in the region of
low to moderate SNR, where multiple transmissions are required to
help correct packets erroneously received after the first round. For
high SNR values, ARQ may not be needed because most packets are correct
after the first transmission. Therefore, we focus our analysis on
the SNR region where BLER values, after the first round, are between
$1$ and $10^{-1}$. In this region, an ARQ protocol is essential
to have reliable communication. Our main goal is to analyze the ISI
cancellation capability and the achieved diversity order for the proposed
turbo combining schemes. We, therefore, evaluate the BLER performance
per ARQ round. We also evaluate the throughput improvement offered
by the proposed schemes. The SNR appearing in all figures is per symbol
per receive antenna. For both schemes, we consider five turbo iterations
for decoding an information block at each transmission. We compare
the resulting performance with the outage probability and the MFB.
Note that for the purpose of fair comparison, the computation of the
outage performance does not take into account the rate distortion
as in (\ref{eq:Outage_eq2}). The MFB curves are obtained for each
transmission assuming perfect ISI cancellation and maximum ratio combining
(MRC) of all time, space, multipath, and delay diversity branches.

\subsection{Analysis}

\begin{figure}[t]
\noindent \begin{centering}
\includegraphics[scale=0.49]{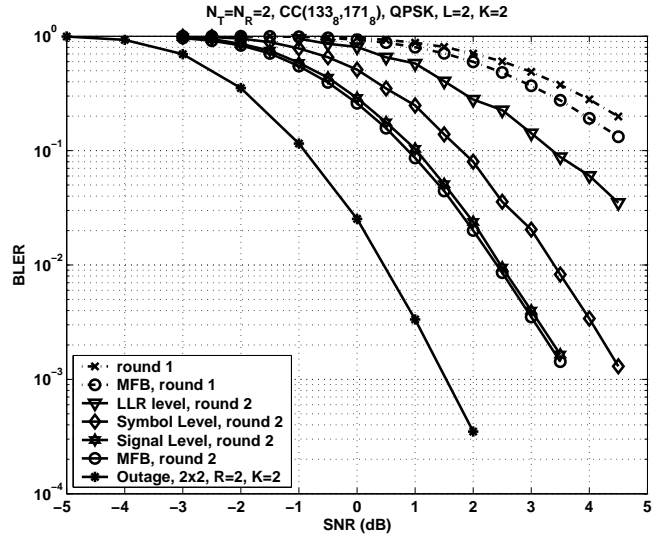}
\par\end{centering}

\caption{\label{fig:BLER_2x2_QPSK}BLER performance comparison for $N_{T}=N_{R}=2$,
CC$\left(133_{8},171_{8}\right)$, QPSK, $K=2$ rounds, and $L=2$
taps.}

\end{figure}

First we consider an ST-BICM code with $N_{T}=2$ and QPSK signaling.
This corresponds to a rate $R=2$. The number of receive antennas
is $N_{R}=2$, and the filter length is $\kappa=9\,\,\left(\kappa_{1}=\kappa_{2}=4\right)$
for all combiners. Fig. \ref{fig:BLER_2x2_QPSK} compares the BLER
performance for the signal-level, symbol-level, and LLR-level combining
with the MFB and the outage probability. For both signal and symbol-level
turbo combining, the performance improvement after the second ARQ
round is very significant compared with LLR-level combining. The signal-level
combining scheme is shown to achieve the MFB while the symbol-level
scheme presents approximately a gap of $1$dB compared with the MFB.
This means that signal-level combining has higher ISI cancellation
capability than symbol-level combining. This result is due to the
fact that in signal-level combining, each ARQ round is considered
as a set of virtual $N_{R}$ receive antennas. This allows the ARQ
delay diversity to be efficiently exploited. On the other hand, both
proposed schemes are shown to achieve the asymptotic slope of the
outage probability.%

\begin{figure}[t]
\noindent \begin{centering}
\includegraphics[scale=0.49]{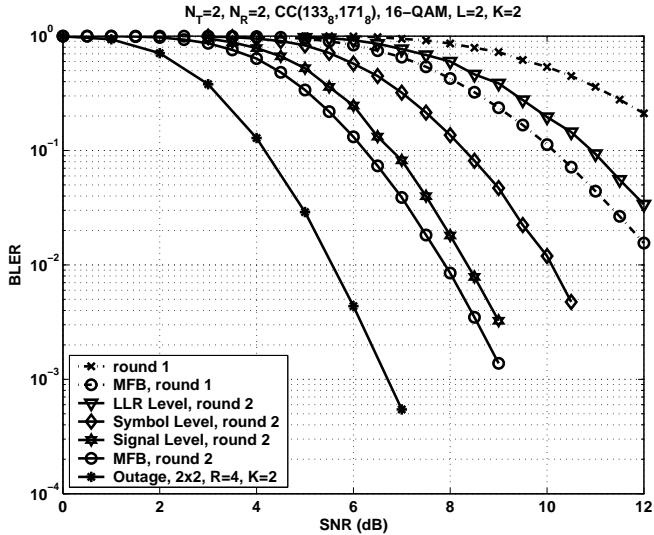}
\par\end{centering}

\caption{\label{fig:BLER_2x2_16QAM}BLER performance comparison for $N_{T}=N_{R}=2$,
CC$\left(133_{8},171_{8}\right)$, 16-QAM, $K=2$ rounds, and $L=2$
taps.}

\end{figure}

\begin{figure}
\noindent \begin{centering}
\includegraphics[scale=0.49]{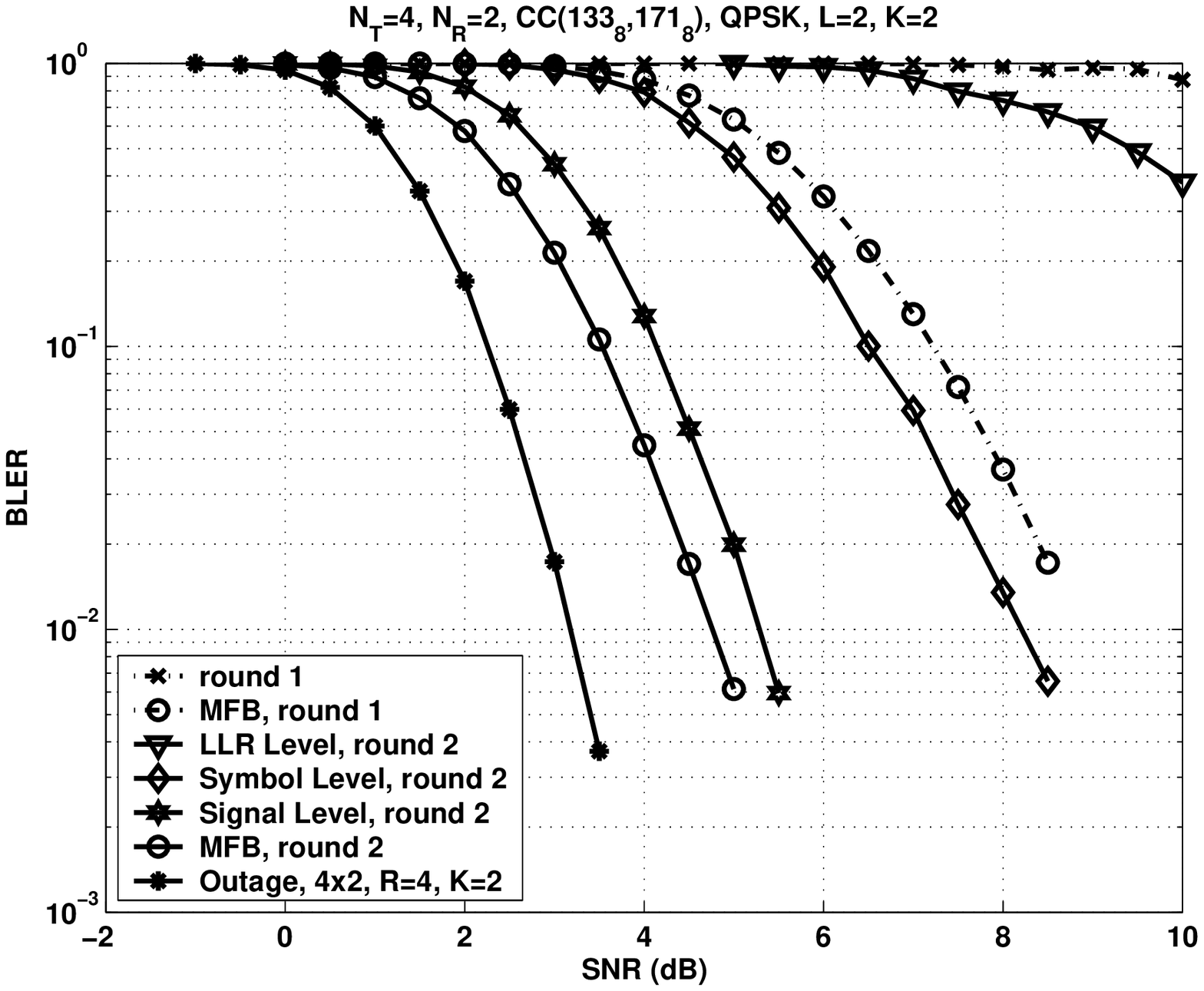}
\par\end{centering}

\caption{\label{fig:BLER_4x2_QPSK}BLER performance comparison for $N_{T}=4,\,\, N_{R}=2$,
CC$\left(133_{8},171_{8}\right)$, QPSK, $K=2$ rounds, and $L=2$
taps.}

\end{figure}

\begin{figure}[t]
\noindent \begin{centering}
\includegraphics[scale=0.49]{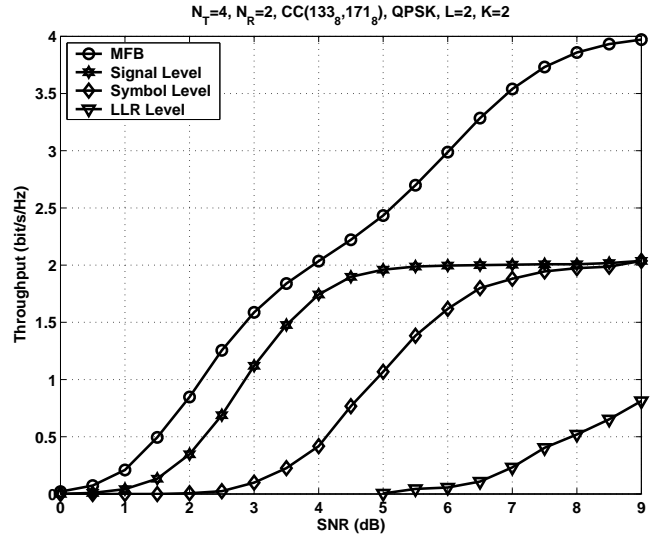}
\par\end{centering}

\caption{\label{fig:Throughput_4x2_QPSK}Throughput performance comparison
for $N_{T}=4,\,\, N_{R}=2$, CC$\left(133_{8},171_{8}\right)$, QPSK,
$K=2$ rounds, and $L=2$ taps.}

\end{figure}

Now, we turn to ST-BICM codes with rate $R=4$. Firstly, we consider
a configuration similar to that of the previous case but using 16-QAM
modulation. The filter length is kept equal to $\kappa=9$. The BLER
performance is reported in Fig. \ref{fig:BLER_2x2_16QAM}. In this
scenario, the signal-level scheme clearly outperforms both the LLR-level
and the symbol-level schemes. Indeed, the gap between the latter and
the MFB is about $2.25$dB. Both proposed techniques asymptotically
achieve the diversity gain of the MIMO ARQ channel. In Fig. \ref{fig:BLER_4x2_QPSK},
we examine a ST-BICM code with $N_{T}=4$, QPSK signaling, and $N_{R}=2$.
Note that this type of {}``unbalanced'' configuration, i.e., more
transmit than receive antennas, is suitable for the forward link.
The filter length is increased to $\kappa=13\,\,\left(\kappa_{1}=\kappa_{2}=6\right)$
for all schemes. The signal-level combining technique is shown to
achieve BLER performance close to the MFB (the gap is less than $0.5$dB),
while both the LLR-level and the symbol-level techniques have a degraded
probability of error (the gap between the symbol-level and the MFB
is more than $3$dB at $2*10^{\text{-2}}$BLER). It is also important
to note that signal-level combining manifests itself in almost achieving
the diversity gain while it is shown that symbol-level combining fails
to do so. This is mainly due to the fact that, at the second ARQ round,
the signal-level scheme constructs a $4\times4$ virtual MIMO-ISI
channel for ISI cancellation and symbol detection, while the MIMO
configuration remains unbalanced in the case of symbol-level combining.
In Fig. \ref{fig:Throughput_4x2_QPSK}, we compare the throughput
performance of the three algorithm for the $4\times2$ configuration.
It is shown that signal-level combining offers higher throughput.
Also, note that while the MFB achieves the maximum throughput of $4$bit/s/Hz,
the proposed techniques saturate around $2$bit/s/Hz because most
of the packets received in the first ARQ round are erroneous. 

Finally, note that in practical systems, channel estimation presents
the bottle-neck that causes performance loss. In \cite{Ait-Idir_WCNC_08},
we evaluated the BLER performance for a low-rate ST-BICM code (typically,
$N_{T}=N_{R}=2$, and $R=2$) with imprecise channel estimates and
using signal-level turbo packet combining. We have shown that when
MMSE channel estimation is performed in a turbo fashion together with
turbo packet combining (i.e., channel is iteratively re-estimated
at each ARQ round using both pilot symbols and soft LLRs), the performance
loss is less than $0.5$dB when $K=2$, and does not exceed $1$dB
when the ARQ delay is increased to $K=3$. Also, we have shown that
even for the case of short-term static dynamic, turbo channel estimation
can offer attractive BLER performance without requiring the re-transmission
of the pilot sequence since channel estimation in subsequent ARQ rounds
can rely only on soft LLRs. %

\section{Conclusion\label{sec:Conclusions}}

In this paper, we considered the design of efficient turbo packet
combining schemes for MIMO ARQ protocols operating over frequency
selective channels. First of all, we derived the structure of the
optimal MAP packet combiner that exploits all the diversities available
in the MIMO-ISI ARQ channel to perform transmission combining. Inspired
by \cite{Caire_IT01,DMD_Gamal_IT_2006}, we then investigated the
outage probability and the outage-based power loss for Chase-type
MIMO ARQ protocols operating over ISI channels. Then, we introduced
two MMSE-based turbo combining schemes that exploit the delay diversity
to perform transmission combining. The signal-level scheme considers
an ARQ round as a set of virtual receive antennas and performs packet
combining jointly with ISI cancellation. The symbol-level scheme separately
equalizes multiple transmissions, while combining is performed at
the level of filter outputs. We showed that both combining schemes
have computational complexities similar to that of the conventional
LLR-level combining. Finally, we presented simulation results that
demonstrated that signal-level combining provides better BLER and
throughput performance than that of symbol-level and LLR-level combining. 

\begin{center}
\textsc{Acknowledgment}
\par\end{center}

The authors would like to thank Prof. Tolga Duman for the comments
he provided about an earlier version of this paper. They also would
like to thank Prof. Angel Lozano for coordinating the review process,
and the three anonymous reviewers for their very helpful comments
and suggestions. 

\vspace{-1cm}
\begin{biography}[{\includegraphics[width=1in,clip,keepaspectratio]{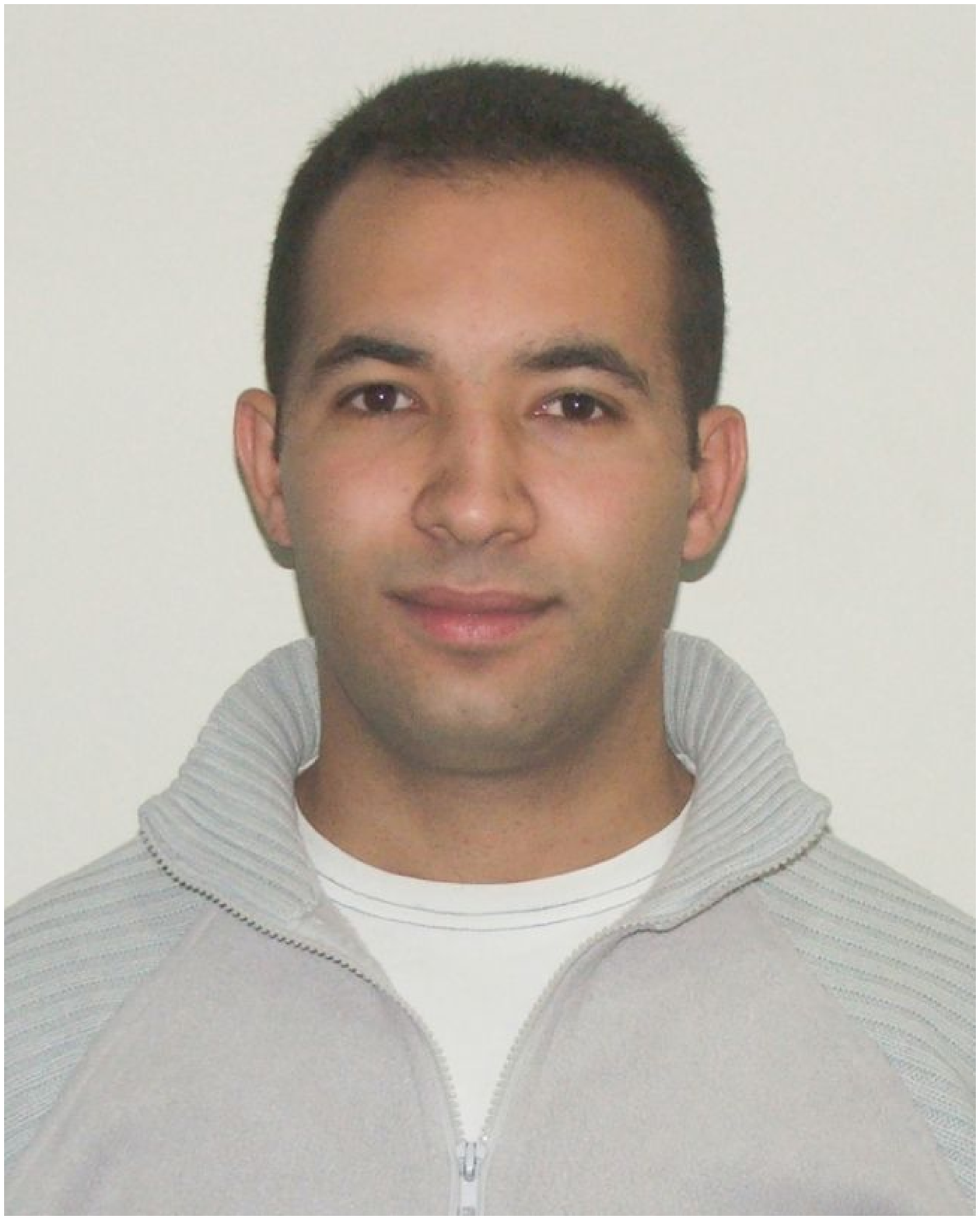}}]{Tarik Ait-Idir}
(S'06--M'07) was born in Rabat, Morocco, in 1978. He received the "diplôme d'ingénieur d'état" in telecommunications from INPT, Rabat, and the Ph.D. degree in electrical engineering from ENST Bretagne, Brest, France, in 2001, and 2006, respectively.\\ \indent He is currently an Assistant Professor of wireless communications at the Communication Systems department, INPT, Rabat. He is also an adjunct researcher with Institut Telecom / Telecom Bretegne/LabSticc. From July 2001 to February 2003 he was with Ericsson. His research interests include PHY and cross-layer aspects of MIMO systems, relay communications, and dynamic spectrum management. \\ \indent Dr. Ait-Idir has been on the technical program committee of several IEEE conferences, including ICC, WCNC, PIMRC, and VTC, and chaired some of their sessions. He has been a technical co-chair of the MIMO Systems Symposium at IWCMC 2009.   
\end{biography}

~

~

~

~

~

~

~

~

~

~~

~

~

~~

~

~

~~

~

~

~~

~

~

~~

~

~

~~

~

~

~~

~

~

~~

~

~

~~

~

~

~~

\noindent \begin{biography}[{\includegraphics[width=1in,clip,keepaspectratio]{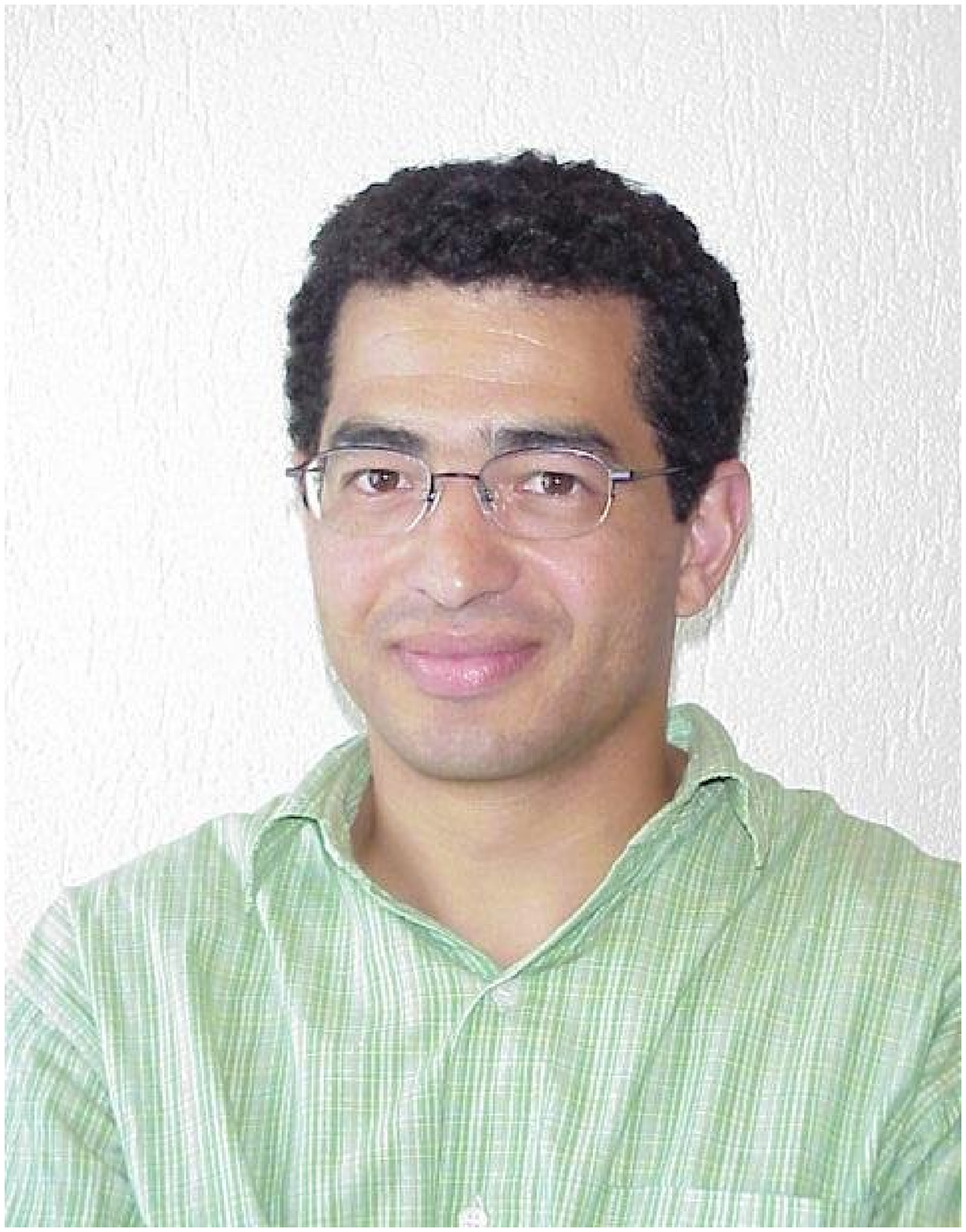}}]{Samir Saoudi}
(M'01) was born in Rabat, Morocco, on November 28, 1963. He received the "diplôme d'ingénieur d'état" from ENST Bretagne, Brest, France, in 1987, the Ph.D. degree in telecommunications from the 'Université de Rennes-I' in 1990, and the "Habilitation à Diriger des Recherches en Sciences" in 1997.\\ \indent Since 1991, he has been with the Signal and Communications department, Institut Telecom / Telecom Bretegne/LabSticc, where he is currently a Professor. He is also with Université Européenne de Bretagne. In summer 2009, he has visited Orange Labs-Tokyo. His research interests include speech and audio coding, non parametric probability density function estimation, CDMA techniques, multiuser detection and MIMO techniques for UMTS and HSPA applications. His teaching interests are signal processing, probability, stochastic processes and speech processing. \\ \indent Dr. Saoudi supervised more than 20 Ph.D. Students. He is the author and/or coauthor of around eighty publications. He has been the general chairman of the second International Symposium on Image/Video Communications over fixed and mobile networks (ISIVC'04).
\end{biography}

\begin{thebibliography}{48}
\bibitem{PeisaEricsson_VTCS07}J. Peisa, S. Wager, M. Sagfors, J.
Torsner, B. Goransson, T. Fulghum, C. Cozzo, and S. Grant, {}``High
speed packet access evolution - concept and technologies,{}`` \emph{in
Proc. 65th IEEE veh. tech. conf. VTC'07 Spring,} Dublin, Ireland,
Apr. 2007.

\bibitem{Woliansky}P. W. Wolniansky, G. J. Foschini, and R. A. Valenzuela,
{}``V-BLAST : an architecture for realizing very high data rates
over the rich scattering wireless channel,\textquotedbl{} \emph{in
Proc. Int. Symp. Signals, Systems, Electron.} , Pisa, Italy, Sep.
1998.

\bibitem{ChaseTComm1985}D. Chase, {}``Code combining--a maximum-likelihood
decoding approach for combining an arbitrary number of noisy packets,''
\emph{IEEE Trans. Commun.,} vol. COM-33, no. 5, pp. 385-393, May 1985.

\bibitem{Harvey_WickerTComm1994}B. Harvey, and S. Wicker, {}``Packet
combining systems based on the Viterbi decoder,'' \emph{IEEE Trans.
Commun.,} vol. 42, no. 2-4, pp. 1544-1557, Feb.-Apr. 1994.

\bibitem{SamraDing_ICASSP_02}H. Samra, and Z. Ding, ``Integrated
iterative equalization for ARQ systems,''\emph{ in Proc.} \emph{IEEE
Int. Conf. on Acoustics, Speech, and Signal Processing (ICASSP)},
Orlando, FL, May 2002.

\bibitem{SamraDing_ISIT02}H. Samra, and Z. Ding, ``Integrated iterative
equalization for ARQ systems,'' \emph{in Proc. IEEE Int. Symp. on
Info. Theory (ISIT)}, Lausanne, Switzerland, Jun., 2002.

\bibitem{SamraDing_Asilomar2002}H. Samra, and Z. Ding, ``Precoded
integrated equalization for packet retransmissions,'' \emph{in Proc.
36th IEEE Asilomar Conf. on Signals, Systems, and Computers,} Monterey,
CA, Nov., 2002.

\bibitem{Samra_DingIEqual_TComm05}H. Samra, and Z. Ding, ''A hybrid
ARQ protocol using integrated channel equalization,'' \emph{IEEE
Trans. Commun.}, vol. 53, no. 12, pp. 1996-2001, Dec. 2005. 

\bibitem{Doan_Narayanan_ItertCombTComm02}D. N. Doan, and K. R. Narayanan,
{}``Iterative packet combining schemes for intersymbol interference
channels,'' \emph{IEEE Trans. Commun.}, vol. 50, no. 4, pp. 560-570,
Apr. 2002.

\bibitem{Dabak_ICC03}E. N. Onggosanusi, A. G. Dabak, Y. Hui, and
G. Jeong, {}``Hybrid ARQ transmission and combining for MIMO systems,''
\emph{in Proc. IEEE Int. Conf. Commun.,} vol. 5, May 2003, pp. 3205-3209.

\bibitem{Samra_Ding04}H. Samra, and Z. Ding, ``Sphere decoding for
retransmission diversity in MIMO flat-fading channels,''\emph{ in
Proc.} \emph{IEEE Int. Conf. on Acoustics, Speech, and Signal Processing
(ICASSP)}, Montreal, Canada, May 2004.

\bibitem{Ding_MIMOARQ_Sphere_SigProc}H. Samra, and Z. Ding, {}``New
MIMO ARQ protocols and joint detection via sphere decoding,'' \emph{IEEE
Trans. Sig. Proc.} vol. 54, no. 2, pp. 473-482, Feb. 2006.

\bibitem{H_Zheng_et_al_PIMRC02}H. Zheng, A. Lozano, and M. Haleem,
{}``Multiple ARQ processes for MIMO systems,'' \emph{in Proc. 13th
IEEE} \emph{Intern. Symp. Personal Indoor and Mobile Radio Commun.
(PIMRC),} Lisbon, Portugal, Sep. 2002.

\bibitem{ZhizhongDing_RiceICC03}Zhihong Ding, and M. Rice, {}``Type-i
hybrid-ARQ using MTCM spatio-temporal vector coding for MIMO systems,''
\emph{in Proc.} \emph{IEEE Int. Conf. on Commun. (ICC), }Anchorage,
AK, May 2003.

\bibitem{Hottinen_et_al_ISSPA03}A. Hottinen, and O. Tirkkonen, {}``Matrix
modulation and adaptive retransmission,'' \emph{in Proc. 7th IEEE}
\emph{Intern. Symp. Sig. Proc. and Applications (ISSPA), }Paris, France,
Jul. 2003.

\bibitem{Koike_ICC04}T. Koike, H. Murata, and S. Yoshida, {}``Hybrid
ARQ scheme suitable for coded MIMO transmission,'' \emph{in Proc.}
\emph{IEEE Int. Conf. on Commun. (ICC), }Paris, France, Jun. 2004.

\bibitem{Ibi_el_al_CommLett_06}S. Ibi, T. Matsumoto, S. Sampei, and
N. Morinaga, \textquotedblleft{}EXIT chart-aided adaptive coding for
MMSE turbo equalization with multilevel BICM,\textquotedblright{}
\emph{IEEE Commun. Lett}., vol. 10, no. 6, pp. 486\textendash{}488,
Jun. 2006.

\bibitem{Krishnaswamy_VTCfall06}D. Krishnaswamy, and S. Kalluri,
\textquotedblleft{}Multi-level weighted combining of retransmitted
vectors in wireless communications,\textquotedblright{} \emph{in Proc.
IEEE Veh. Technol. Conf., (VTC)}, Montreal, Canada, Sep. 2006.

\bibitem{Jang_et_al_ISIT07}E. W. Jang, J. Lee, H.-L. Lou, and J.
M. Cioffi, \textquotedbl{}Optimal combining schemes for MIMO systems
with hybrid ARQ,\textquotedbl{} \emph{in Proc. IEEE Intern. Symp.
Info. Theory (ISIT),} Nice, France, Jun. 2007.

\bibitem{Ibi_el_al_TVT_07}S. Ibi, T. Matsumoto, R. Thoma, S. Sampei,
and N. Morinaga, {}``EXIT chart-aided adaptive coding for multilevel
BICM with turbo equalization in frequency-selective MIMO channels,''
\emph{IEEE Trans. Veh. Technol.,} vol. 56, no. 6, pp. 3757\textendash{}3769,
Nov. 2007.

\bibitem{Garg_Adachi_JSAC06}D. Garg, and F. Adachi, {}``Packet access
using DS-CDMA with frequency-domain equalization,'' \emph{IEEE Journal
Select. Areas in Commun.,} vol. 24, no. 1, Jan. 2006.

\bibitem{AdachiVTC2005}A. Nakajima, D. Garg, and F. Adachi, {}``Throughput
of turbo coded hybrid ARQ using single-carrier MIMO multiplexing,''
\emph{in Proc. 61st IEEE veh. technol. conf. VTC'05 Spring,} Stockholm,
Sweden, 2005.

\bibitem{Adachi_HARQ_SCMIMO_VTC06}A. Nakajima, and F. Adachi, {}``Iterative
joint PIC and 2D MMSE-FDE for turbo-coded HARQ with SC-MIMO multiplexing,''
\emph{in Proc. 63rd IEEE veh. technol. conf. VTC'06 Spring,} pp. 2503-2507,
Melbourne, Australia, May. 2006.

\bibitem{DMD_Gamal_IT_2006}H. El Gamal, G. Caire, and M. O. Damen,
{}``The MIMO ARQ channel: diversity\textendash{}multiplexing\textendash{}delay
tradeoff,'' \emph{IEEE Trans. Inf., Theory,} vol. 52, no. 8, Aug.
2006, pp. 3601-3621. 

\bibitem{Zheng-Tse}L. Zheng, and D. N. C. Tse, \textquotedblleft{}Diversity
and multiplexing:Afundamental tradeoff in multiple antenna channels,\textquotedblright{}
\emph{IEEE Trans. Inf. Theory,} vol. 49, no. 5, pp. 1073\textendash{}1096,
May 2003.

\bibitem{Medles-Slock_ISIT05}A. Medles, and D. T. M. Slock, {}``Optimal
diversity vs multiplexing tradeoff for frequency selective MIMO channels'',
\emph{in Proc Intern. Symp. Info. Theory} \emph{(ISIT),} Adelaide,
Australia, Sept. 2005. 

\bibitem{Slock_ITW2007}D. T. M. Slock, ''On the diversity-multiplexing
tradeoff for frequency-selective MIMO channels,'' \emph{in Proc.
Info. Theory and App. (ITA) Workshop}, San Diego, CA, Jan-Feb., 2007.

\bibitem{Bolskei_ISIT07}P. Coronel, and H, Bölcskei, {}``Diversity-multiplexing
tradeoff in selective-fading MIMO channels,'' \emph{in Proc. Intern.
Symp. Info. Theory} \emph{(ISIT),} Nice, France, Jun. 2007. 

\bibitem{Holliday_Glodsmith_Poor_ICC06}T. Holliday, A. Goldsmith,
and H. V. Poor, {}``The impact of delay on the diversity, multiplexing,
and ARQ tradeoff,'' \emph{in Proc.} \emph{IEEE Int. Conf. on Commun.
(ICC), }Istanbul, Turkey, Jun. 2006.

\bibitem{Chuang_et_al_DMD_IT08}A. Chuang, A. Guillen i Fabregas,
L.K. Rasmussen, I.B. Collings, \textquotedblleft{}Optimal throughput-diversity-delay
tradeoff in MIMO ARQ block-fading channels,\textquotedblright{} \emph{IEEE
Trans. Inf., Theory,} vol. 54, no. 9, Sep. 2008, pp. 3968-3986.

\bibitem{Ariyavisitakul_ICC00}S. L. Ariyavisitakul, {}``Turbo space-time
processing to improve wireless channel capacity,'' \emph{in Proc.
IEEE Commun. Conf.,} vol. 3, pp. 1238-1242, Jun. 2000.

\bibitem{Tonello}A. M. Tonello, {}``MIMO MAP equalization and turbo
decoding in interleaved space time coded systems'', \emph{IEEE Trans.
Commun.}, vol. 51, no. 2, pp. 155-160, Feb. 2003.

\bibitem{Visoz_TComm03}R. Visoz, and A. O. Berthet, {}``Iterative
decoding and channel estimation for space-time BICM over MIMO block
fading multipath AWGN channel'', \emph{IEEE Trans. Commun.,} vol.
51, no. 8, pp. 1358-1367, Aug. 2003.

\bibitem{vandendorpe_SigProc_04}X. Wautelet, A. Dejonghe, and L.
Vandendorpe, {}``MMSE-based fractional turbo receiver for space-time
BICM over frequency selective MIMO fading channels,'' \emph{IEEE
Trans. Sig. Proc.,} vol. SIG-52, pp. 1804-1809, Jun. 2004.

\bibitem{Visoz-groupMMSE-journal}R. Visoz, A. O. Berthet, and S.Chtourou,
\textquotedbl{}A new class of iterative equalizers for space-time
BICM over MIMO block fading multipath AWGN channel,\textquotedbl{}
\emph{IEEE Trans. Commun.}, vol. 53, no. 12, pp. 2076-2091, Dec. 2005. 

\bibitem{Ait-idir_TVT}T. Ait-Idir, S. Saoudi, and N. Naja, {}``Space-time
turbo equalization with successive interference cancellation for frequency
selective MIMO channels,'' \emph{IEEE Trans. Veh. Technol., }vol.
57, no. 5, pp. 2766-2778, Sep. 2008.

\bibitem{Tuchler_TSP_2002}M. Tüchler, A. C. Singer, and R. Koetter,
{}``Minimum mean squared error equalization using \emph{a priori}
information,'' \emph{IEEE Trans. Sig. Proc,} vol. 50, no. 3, pp.
673\textendash{}683, Mar. 2002.

\bibitem{Ait-Idir_WCNC_08}T. Ait-Idir, H. Chafnaji, and S. Saoudi,
\textquotedbl{}Joint hybrid ARQ and Iterative Space-Time Equalization
for Coded Transmission over the MIMO-ISI Channel,\textquotedbl{} \emph{in
Proc. IEEE Wireless Commun. Net. Conf. (WCNC),} Las Vegas, NV, Mar-Apr.
2008.

\bibitem{Shamai_Ozarow_WynerIT94}L. H. Ozarow, S. Shamai, and A.
D. Wyner, \textquotedblleft{}Information theoretic considerations
for cellular mobile radio,\textquotedblright{} \emph{IEEE Trans. Inform.
Theory,} vol. 43, pp. 359\textendash{}378, May 1994.

\bibitem{Telatar_EurpTelecom_99}I. E. Telatar, \textquotedblleft{}Capacity
of multi-antenna Gaussian channels,\textquotedblright{} \emph{Europ.
Trans. Telecommun.}, vol. 10, no. 6, pp. 585\textendash{}595, Nov./Dec.1999.

\bibitem{Foshini_Gans_WPC_98}G. J. Foschini and M. J. Gans, \textquotedblleft{}On
limits of wireless communications in a fading environment when using
multiple antennas,\textquotedblright{} Wireless Personal Commun.,
vol. 6, pp. 311\textendash{}335, Mar. 1998.

\bibitem{Tse_Viswanath_Book}D. Tse, and P. Viswanath, {}``Fundamentals
of Wireless Communication,'' Cambridge University Press, May 2005.

\bibitem{ElGamal_IT04}H. El Gamal, A. R. Hammons, Y. Liu, M. P. Fitz,
and O. Y. Takeshita, {}``On the design of space\textendash{}time
and space\textendash{}frequency codes for MIMO frequency-selective
fading channels,'' \emph{IEEE Trans. Inform. Theory,} vol. 49, no
9, pp. 2277\textendash{}2292, Sep. 2003.

\bibitem{Duman_Tcomm04}Z. Zhang, T. M. Duman, and E. M. Kurtas, {}``Achievable
information rates and coding for MIMO systems over ISI channels and
frequency-selective fading channels,'' \emph{IEEE Trans. Commun.,}
vol. 52, no 10, pp. 1698\textendash{}1710, Oct. 2004.

\bibitem{RenewalTheory_Wolff}R. Wolff, \emph{Stochastic Modeling
and the Theory of Queues.} Upper Saddle River, NJ: Prentice-Hall,
1989.

\bibitem{RenewalTheory_Zorzi}M. Zorzi, and R. R. Rao, \textquotedblleft{}On
the use of renewal theory in the analysis of ARQ protocols,\textquotedblright{}
\emph{IEEE Trans. Commun.,} vol. 44, pp. 1077\textendash{}1081, Sept.
1996.

\bibitem{Caire_IT01}G. Caire, and D. Tuninetti, \textquotedblleft{}ARQ
protocols for the Gaussian collision channel,\textquotedblright{}
\emph{IEEE Trans. Inf. Theory,} vol. 47, no. 4, pp. 1971\textendash{}1988,
Jul. 2001.

\bibitem{Haykin}S. Haykin, \emph{Adaptive Filter Theory,} 3rd Ed.
Upper Saddle River, NJ: Prentice-Hall, 1996.

\end{thebibliography}
\end{document}